\documentclass[prd,aps,preprint,groupedaddress,superscriptaddress,nofootinbib]{revtex4}
\usepackage{graphicx}
\usepackage{amsmath,amssymb,bm}
\usepackage{hyperref}
\usepackage{autobreak,mathtools}
\usepackage{comment}
\usepackage[dvipsnames, usenames]{xcolor}
\usepackage[scr=boondoxo,scrscaled=1.05]{mathalfa}
\usepackage{pifont}
\usepackage[normalem]{ulem}

 \usepackage{array, booktabs, pifont, rotating}
\definecolor{p-r}{RGB}{171, 40, 52}
\hypersetup{colorlinks=true, citecolor=p-r, linkcolor=p-r,urlcolor=p-r}

\begin{document}

\title{Regularizing gravity in the presence of matter fields}
\author{Ra\'ul Carballo-Rubio}
\affiliation{Instituto de Astrof\'isica de Andaluc\'ia (IAA-CSIC),
Glorieta de la Astronom\'ia, 18008 Granada, Spain}
\author{Chiara Coviello}
\affiliation{Department of Physics, King’s College London, The Strand, London, WC2R 2LS, UK}
\author{Vania Vellucci}
\affiliation{Quantum Theory Center (${\hbar}$QTC) \& D-IAS, IMADA at Southern Denmark Univ., Campusvej 55, 5230 Odense M, Denmark}

\begin{abstract}
We investigate how matter fields affect spacetime regularity in a class of gravitational theories with second-order field equations in spherical symmetry, of which general relativity is a special case, and which may encode effective corrections from an underlying quantum theory of gravity. Focusing on spherically symmetric black holes and cosmological spacetimes in theories in which vacuum spacetimes display bounded curvature invariants, we derive the conditions under which regular geometries can be supported by static and perturbative scalar fields, electromagnetic fields, and radiation fluids. We discuss the impact of matter on the regularity requirements of curvature boundedness and geodesic completeness relative to vacuum, finding in general that coupling to  matter does not eliminate the existence of theories with regular solutions.
\end{abstract}

\maketitle

\section{Introduction}

It is generally expected that spacetime singularities inside black holes will be resolved within a complete theory of gravity, possibly by incorporating quantum effects. Although current approaches to quantum gravity are still far from offering a fully developed framework, their leading low-energy consequences can often be captured through effective modifications of general relativity (see, for example, recent reviews on quantum gravitational collapse in~\cite{Eichhorn:2022bgu,Platania:2023srt,Ashtekar:2023cod}). Nevertheless, progress in this direction has faced significant obstacles. In particular, there are indications that achieving regular black hole interiors requires field equations involving derivatives beyond second order~\cite{Knorr:2022kqp,Borissova:2023kzq}. Such higher-order field equations, however, introduce serious conceptual and technical difficulties in dynamical regimes (see, e.g.,~\cite{Cayuso:2020lca,Cayuso:2023xbc}).

On the other hand, it is well known that singularities inside black hole interiors can be removed at the kinematical (geometric) level, leading to the concept of regular black hole initially studied by Bardeen, Frolov, Vilkovisky, Dymnikova, Hayward and others~\cite{Bardeen1968,Frolov:1981mz,Dymnikova:1992ux,Hayward:2005gi,Frolov:2014jva,Dymnikova:2015yma}. Due to being the best-motivated alternative to (singular) classical black holes~\cite{Carballo-Rubio:2025fnc}, regular black holes have been extensively studied, from formal~\cite{Carballo-Rubio:2019nel,Carballo-Rubio:2019fnb,Carballo-Rubio:2021wjq,Mazza:2021rgq,Franzin:2022wai,Carballo-Rubio:2022nuj,LimaJunior:2025uyj,Eichhorn:2025pgy} to phenomenological~\cite{Cardoso:2019rvt,Carballo-Rubio:2018jzw,Eichhorn:2021etc,Eichhorn:2022oma,Eichhorn:2022bbn,Franzin:2022iai,Cardoso:2023guh,Mazza:2023mvx} perspectives. However, the lack of known dynamical laws leading to the formation of such spacetimes has been a significant challenge for the solidification of regular black holes as self-consistent and predictive alternative models~\cite{Carballo-Rubio:2018pmi}.

A significant step forward in the consolidation of regular black holes is the proof that, in spherical symmetry in four or more spacetime dimensions, their dynamical properties can be described by using second-order gravitational field equations in which the Einstein tensor is deformed into a conserved tensor constructed from up to second-order derivatives of the metric~\cite{Carballo-Rubio:2025ntd}. This description in terms of the so-called master field equations is valid for any regular black hole metric, with static metrics arising as vacuum solutions~\cite{Ziprick:2010vb,Kunstatter:2015vxa} while collapsing metrics can be constructed by using  the same matter contents as in general relativity~\cite{Boyanov:2025pes}. With the treatment being valid in four and higher dimensions, it provides a unifying framework in which to include earlier work on non-polynomial gravities~\cite{Colleaux:2017ibe,Colleaux:2019ckh} (and more recent studies~\cite{Bueno:2025zaj,Borissova:2026wmn,Borissova:2026krh}) as well as on the higher-dimensional theories known as quasitopological gravities~\cite{BuenoRegularBlack2025c,Bueno:2024eig,Bueno:2024zsx,Bueno:2025gjg,Hennigar:2025yqm,PinedoSoto:2026hfm,2026arXiv260310110B}. Diverse applications of these master field equations for spherically symmetric gravitational fields can be found in~\cite{Borissova:2026dlz,Borissova:2026wmn,Borissova:2026krh,Borissova:2026klg,Arrechea:2026ngi,Borissova$g_ttg_rr1$2026,Carballo-Rubio:2026mvj,Mazza:2026taj,BorissovaFormationExtremal2026,Thaalba:2026abz,Carballo-Rubio:2026mns}. A complementary treatment from a Hamiltonian perspective has been developed in~\cite{Alonso-BardajiDynamicalTheory2025,Zhang:2025ccx,Zhang:2026tjc}.

In this work we investigate how specific deformations of general relativity, in which static vacuum solutions are regular, behave in the presence of matter fields, in particular scalar fields, electromagnetic fields and perfect fluids in a cosmological setting. The paper is organized as follows, as a follow-up of the discussion in the shorter paper~\cite{Carballo-Rubio:2026mvj}. We briefly review the field equations and families of theories we will be working with in Sec.~\ref{sec:vacuum}. We then discuss solutions of these equations in the presence of matter fields: scalar fields in Sec.~\ref{sec:scalar}, electromagnetic fields in Sec.~\ref{sec:maxwell}, and perfect fluids in a cosmological context in Sec.~\ref{sec:cosm}. We wrap up the paper discussing the aspects shared by all these different situations and drawing general conclusions in Sec.~\ref{sec:disc}, while additional technical details are presented for completeness in Apps.~\ref{app:perturbative} and~\ref{sec:K_nonzero}.
 
\section{Review of the master field equations for spherically symmetric gravitational fields}\label{sec:vacuum}

%%%%%%%%%%%%%%%%%%%%
\subsection{Field equations}
%%%%%%%%%%%%%%%%%%%%

The most general field equations for spherically symmetric gravitational fields, in which the Einstein tensor is deformed into a conserved tensor constructed from up to second-order derivatives of the metric, are given by~\cite{Carballo-Rubio:2025ntd}:
\begin{equation}\label{eq:sphnew}
\mathscr{G}_{\mu\nu}(q,r)=\frac{1}{r^2}\mathscr{E}_{ab}\delta^a_\mu\delta^b_\nu-\frac{1}{4}r\mathscr{F}\gamma_{ij}\delta^i_\mu\delta^j_\nu=8\pi T_{\mu\nu},
\end{equation}
where
\begin{align}\label{eq:geneqs1}
\mathscr{E}_{ab}(q,r)&=\bm{\beta}\nabla_a\nabla_b r-q_{ab}\left(\frac{1}{2}\bm{\alpha}+\bm{\beta}\square r \right)+\left(\partial_\chi\bm{\alpha}-\partial_r\bm{\beta} \right)\nabla_ar\nabla_br
,\\
\mathscr{F}(q,r)&=-\bm{\beta}\mathcal{R}+2\partial_r\bm{\beta}\square r+\partial_r\bm{\alpha}+2\partial_\chi\bm{\beta}\left(\left(\square r\right)^2-\nabla_a\nabla_b r\nabla^a\nabla^br\right)\nonumber\\
&-2\partial_r\left(\partial_\chi\bm{\alpha}-\partial_r\bm{\beta}\right)\chi-2\left(\partial_\chi\bm{\alpha}-\partial_r\bm{\beta} \right)\square r-2\partial_\chi\left(\partial_\chi\bm{\alpha}-\partial_r\bm{\beta}\right)\nabla_a r\nabla^a\chi,\label{eq:geneqs2}
\end{align}
Here, we define $\chi=\nabla_a r\nabla^a r$, where $r(x)$ is the dilaton field in a metric ansatz of the ``warped product'' form~\cite{ONeillbook}:
\begin{equation}\label{eq:symred}
g_{\mu\nu}(y)\text{d}y^\mu\text{d}y^\nu=q_{ab}(x)\text{d}x^a\text{d}x^b+r^2(x)\gamma_{ij}\text{d}\theta^i\text{d}\theta^j.
\end{equation}
In these expressions, $q_{ab}(x)$ represents a 2-dimensional metric and $\mathcal{R}$ its associated Ricci scalar.

While the occurrence of terms of the form $(\square r)^2$ and $\nabla_a\nabla_br\nabla^a\nabla^br$ in Eq.~\eqref{eq:geneqs2} may lead to well-posedness issues, taking the trace of Eq.~\eqref{eq:geneqs1} yields expressions for $\square r$ and $\nabla_a\nabla_b r$ in terms of first-order derivatives of $r$. Hence, only ${\cal R}$ contributes with second derivatives of $q_{ab}$ to the left-hand side of Eq.~\eqref{eq:geneqs2}. This indicates that the system can be suitable for defining a well-posed problem for appropriate coordinate conditions.

The spherically symmetric Einstein field equations are recovered as the particular case
\begin{equation}\label{eq:grchoicesalphabeta}
\bm{\alpha}^{\rm(GR)}=-2(\chi-1),\quad \bm{\beta}^{\rm(GR)}=-2r,
\end{equation}
for which Eqs.~(\ref{eq:geneqs1}-\ref{eq:geneqs2}) are greatly simplified. In particular, the last two terms in the first line of Eq.~\eqref{eq:geneqs2} cancel because $\partial_r\bm{\alpha}^{\rm(GR)}=0$ and $\partial_\chi\bm{\beta}^{\rm(GR)}=0$, while the last term in Eq.~\eqref{eq:geneqs1}, as well as all the terms in the second line of Eq.~\eqref{eq:geneqs2} vanish due to $\partial_\chi\bm{\alpha}^{\rm(GR)}=\partial_r\bm{\beta}^{\rm(GR)}$.

Additional considerations are needed to guarantee that a meaningful 4-dimensional description can be reconstructed from this effectively 2-dimensional description~\cite{Carballo-Rubio:2025ntd}. The spherical coordinates used to write the metric in Eq.~\eqref{eq:symred} display a coordinate singularity at $r=0$. In order to guarantee physical regularity, we follow the same standard procedure as in the treatment of symmetric spacetimes in general relativity (see, e.g.,~\cite{Rinne:2005sk,Rinne:2005df} for a discussion of the axisymmetric case) and demand the existence of locally Cartesian coordinates around the center of symmetry, in which the metric coefficients can be expanded in Taylor series and are thus analytic (this assumption can be weakened to differentiability up to a specific order, although we will not consider such cases here). This imposes additional constraints on the metric coefficients around $r=0$, which are more easily written explicitly choosing the admissible coordinate choice $x^1=r$:
\begin{equation}\label{eq:regtay}
q_{00}=-h(t,r^2),\quad q_{0r}=rj(t,r^2),\quad q_{rr}=f(t,r^2).
\end{equation}
These conditions being satisfied discriminates between the space of 2-dimensional fields $q_{ab}(x)$, $r(x)$ that can be understood as the building blocks of a four-dimensional metric~\eqref{eq:symred}, and the complementary space of these 2-dimensional fields for which this four-dimensional interpretation is not viable. Equivalently, these can be understood as conditions that determine whether geodesic completeness at $r=0$ holds~\cite{Zhou:2022yio,Arrechea:2026ngi}.

A direct calculation of the components of the inverse metric $q^{ab}$ and $\chi=q^{rr}$ shows that, as a consequence of~\eqref{eq:regtay},
\begin{equation}
\chi=\chi(t,r^2).
\end{equation}
The same considerations as for the metric tensor are applicable to the tensor $\mathscr{G}_{\mu\nu}(q,r)$ defined in Eq.~\eqref{eq:sphnew}, resulting in the following constraints on the functions $\bm{\alpha}$ and $\bm{\beta}$ (see also the discussions in~\cite{Arrechea:2026ngi,Thaalba:2026abz}):
\begin{equation}\label{eq:refalphbet}
\bm{\alpha}[r,\chi(t,r^2)]=\bm{a}[r^2,\chi(t,r^2)],\quad \bm{\beta}[r,\chi(t,r^2)]=r\bm{b}[r^2,\chi(t,r^2)].
\end{equation}
It can be easily checked that these relations are satisfied by the expressions valid for general relativity, that were provided in Eq.~\eqref{eq:grchoicesalphabeta}.

%%%%%%%%%%%%%%%%%%%%
\subsection{Vacuum solutions}
%%%%%%%%%%%%%%%%%%%%
\label{sec:vacuumsols}

The deformation of the Einstein tensor in terms of $\bm{\alpha}$ and $\bm{\beta}$ in Eq.~\eqref{eq:sphnew} induces a corresponding deformation in the vacuum solutions. In general, we can write any spherically symmetric metrics, without loss of generality, as
\begin{equation}
    \text{d}s^2= -h(t,r)\text{d}t^2+  \frac{\text{d}r^2}{f(t,r)}+ r^2 \text{d}\Omega^2,
    \label{eq:metric_back}
\end{equation}
for which $\chi=f(t,r)$. The vacuum field equations resulting from considering $T_{ab}=0$ in Eq.~\eqref{eq:sphnew} imply that the time dependence drops and that the functions $h(r)$ and $f(r)$ satisfy the equations~\cite{Carballo-Rubio:2025ntd}
\begin{equation}\label{eq:vaceqs}
\frac{\text{d}f(r)}{\text{d}r}=-\frac{\bm{\alpha}\left[r,f(r)\right]}{\bm{\beta}\left[r,f(r)\right]},\quad h(r)=f(r)e^{2\int\text{d}r\left.\frac{\left(\partial_\chi\bm{\alpha}-\partial_r\bm{\beta}\right)}{\bm{\beta}}\right|_{\chi=f(r)}}.
\end{equation}

These equations can be solved order by order by considering the general expansions
\begin{align}
f &=f_0+\sum_{k=n}^\infty f_{k} r^{k}= f_0 + f_{n}\, r^{n} + \dots\,,\label{eq:fexp0}\\
h&= h_0+\sum_{l=m}^\infty h_{l} r^{l}=  h_0 + h_{m}\, r^{m} +\dots\,, \label{eq:hexp0}
\end{align}
with $f_n\neq0$, $h_m\neq0$ and $f_0$, $h_0$ real constants, while $n$ and $m$ are arbitrary integers.

The constraints in Eq.~\eqref{eq:regtay} require that all odd powers in Eqs.~\eqref{eq:fexp0} and~\eqref{eq:hexp0}ee vanish, which in particular implies that $n$ and $m$ must be even numbers. On the other hand, the Kretschmann scalar is bounded at $r\to 0$ if and only if (see, e.g.,~\cite{Carballo-Rubio:2019fnb})
\begin{equation}\label{eq:regconds}
f_0=1,\quad f_1=h_1=0.
%f(0)=1,\quad f'(0)=h'(0)=0.    
\end{equation}

For static solutions, the regularity of the Kretschmann invariant guarantees the regularity of all other (non-derivative) curvature invariants~\cite{Lobo:2020ffi,Carballo-Rubio:2025fhq}. However, this is no longer the case in time-dependent situations. In this paper, we will focus on the regularity of the Kretschmann invariant as a necessary condition for the regularity of (non-derivative) curvature invariants.

For the rest of the paper, when dealing with static solutions we will consider expansions of the form in Eqs.~\eqref{eq:fexp0} and~\eqref{eq:hexp0} with $f_0=1$, namely
\begin{align}
f &=1+\sum_{k=n}^\infty f_{k} r^{k}= 1 + f_{n}\, r^{n} + \dots\,,\label{eq:fexp}\\
h&= h_0+\sum_{l=m}^\infty h_{l} r^{l}=  h_0 + h_{m}\, r^{m} +\dots\,. \label{eq:hexp}
\end{align}
These are general enough to describe the singular Schwarzschild solution as well as regular solutions.

%%%%%%%%%%%%%%%%%%%%
\subsection{Spaces of theories}
%%%%%%%%%%%%%%%%%%%%
\label{sec:spaces}

For the sake of simplicity, we will further restrict the space of theories considered in our analysis. We will assume in all cases the integrability condition~\cite{Carballo-Rubio:2025ntd,Boyanov:2025pes}
\begin{equation}\label{eq:intcond}
\partial_\chi\bm{\alpha}-\partial_r\bm{\beta}=0. 
\end{equation}
This condition restricts vacuum solutions to depend on a single metric function $f(r)=h(r)$, as the expression for $h(r)$ in Eq.~\eqref{eq:vaceqs} shows explicitly (considering situations in which the condition above is not satisfied can be interesting for specific applications, see for instance~\cite{Borissova:2026dlz,Mazza:2026taj}). The remaining function $f(r)$ can be then found by solving an algebraic relation that is obtained by integrating Eq.~\eqref{eq:vaceqs}:
\begin{equation}\label{eq:omegasol}
\left.\bm{\Omega}(r,\chi)\right|_{\chi=f}=4M,    
\end{equation}
where we have normalized this function so that $M$ is the ADM mass, and
\begin{equation}\label{eq:alphabetaomegadef}
\frac{\partial\bm{\Omega}}{\partial r}=\bm{\alpha},\quad \frac{\partial\bm{\Omega}}{\partial \chi}=\bm{\beta}.
\end{equation}
The existence of the potential function $\bm{\Omega}(r,\chi)$ is guaranteed by the condition~\eqref{eq:intcond}. For general relativity, this potential function takes the form
\begin{equation}
\bm{\Omega}^{\rm(GR)}=-2\left(\chi-1\right)r.    
\end{equation}
We will also restrict our attention to theories in the Ziprick--Kunstatter--Maeda--Taves (ZKMT) class of theories, as defined in~\cite{Boyanov:2025pes}:
\begin{equation}\label{eq:zkmtdef}
\bm{\Omega}=2r^3\psi\bm{\Sigma}(\psi)\bm{\Phi}(r),
\end{equation}
where we have defined $\psi=(1-\chi)/r^2$ and taken into account the required behavior of the functions $\bm{\alpha}$ and $\bm{\beta}$ in Eq.~\eqref{eq:refalphbet}. In the parametrization above, general relativity corresponds to $\bm{\Sigma}=\bm{\Phi}=1$.

We will be focusing on two separate subsets of theories. We will work in different ways with each of these subsets both due to their different defining peculiarities but also due to their different behavior observed in the presence of non-gravitational fields~\cite{Carballo-Rubio:2026mvj}, which we aim to understand here in more detail:
\begin{itemize}
    \item The Ziprick--Kunstatter (ZK) family is defined by the condition $\bm{\Sigma}=1$ and includes theories in which vacuum solutions take similar forms to the Bardeen spacetime~\cite{Bardeen:1968xqk,Ziprick:2010vb,Boyanov:2025pes}. Charged solutions in these theories can become singular at $r=0$, even if uncharged vacuum solutions have bounded curvature invariants, unless additional conditions are met~\cite{Carballo-Rubio:2026mvj}. Hence, when analyzing these theories in Secs.~\ref{sec:scalar}, \ref{sec:maxwell} and~\ref{sec:gensta}, we will mostly focus on the behavior of solutions around $r=0$ and make use of perturbative expansions, which we introduce in Sec.~\ref{sec:zkexp} below.
    \item The Kunstatter--Maeda--Taves (KMT) family is defined by $\bm{\Phi}=1$ and includes theories in which vacuum solutions take similar forms to the Hayward spacetime~\cite{Hayward:2005gi,Kunstatter:2015vxa,Boyanov:2025pes}. Charged solutions in these theories can become singular at $r>0$, even if uncharged vacuum solutions have bounded curvature invariants, unless additional conditions are met~\cite{Carballo-Rubio:2026mvj}, but remain regular at $r=0$. Perturbative expansions around $r=0$ are therefore less useful to identify possible pathologies, hence when analyzing these theories in Sec.~\ref{sec:cosm} we will consider matter contents for which the corresponding equations can be suitably studied without recurring to a perturbative expansion, in particular homogeneous and isotropic perfect fluids.
\end{itemize}
%

%%%%%%%%%%%%%%%%%%%%
\subsection{Series expansion of vacuum solutions in the Ziprick--Kunstatter family}\label{sec:zkexp}
%%%%%%%%%%%%%%%%%%%%

As discussed in the section above, when working with theories in the ZK family that satisfy the integrability condition in Eq.~\eqref{eq:intcond} we will mostly analyze perturbative expansions around $r=0$. Hence, it will be useful for the analysis below to write down on-shell expansions of the functions $\bm{\alpha}[r,f(t,r)]$ and $\bm{\beta}[r,f(t,r)]$ in the $r\rightarrow 0$ limit, which remain structurally the same regardless of the matter content and we thus collect here before considering specific sources. We start by writing
\begin{equation}\label{eq:taylor}
    \bm \alpha= r^{p}\sum_{i=0} \alpha_{i} r^{i},  \qquad{}
    \bm \beta=  r^{q}\sum_{j=0} \beta_{j} r^{j}, 
\end{equation}
where $\alpha_0\neq 0$, $\beta_0\neq 0$ and $p,\, q$ are parameters to be determined. These expansions are consistent with Eq.~\eqref{eq:refalphbet} only if $p$ is an even integer and $q$ is an odd integer, which we will keep in mind for later reference.

The expansions of $\bm{\alpha}$ and $\bm{\beta}$ are not independent for theories in the ZK class, in which the following relation is satisfied:
\begin{equation}\label{eq:alphaZK}
\bm{\alpha}=\left(\chi-1\right)\frac{\text{d}\bm{\beta}}{\text{d}r}.
\end{equation}
At leading order, this constraint yields:
\begin{equation}\label{eq:feqpert}
\alpha_0\, r^{p}-q\,\beta_0\, f_{n}\; r^{n+q-1}=0\,.
\end{equation}
Matching the exponents gives
\begin{equation}\label{eq:exp2}
p=n+q-1,   
\end{equation}
while matching the coefficients provides another relation
\begin{equation}\label{eq:coeff2}
\alpha_0 -q\,\beta_0\, f_{n}=0\,.
\end{equation}
The latter relation determines $\alpha_0$ as a function of $\beta_0$ and $q$. On the other hand, note that Eq.~\eqref{eq:exp2} implies that $p-q\geq 1$ for theories with bounded Kretschmann invariant ($n\geq 2$).

The coefficient $q$ is fixed for a given theory, and according to Eq.~\eqref{eq:refalphbet} it must be odd for a four-dimensional interpretation to exist, while the value that $p$ (alternatively, $n$) takes is solution-dependent. In the sections below, we will be solving the field equations for different matter contents and obtain relations between the values of these coefficients. 

For the theories satisfying the integrability condition in Eq.~\eqref{eq:intcond}, in vacuum there is only one remaining independent field equation, as it is for instance clear from the second relation in Eq.~\eqref{eq:vaceqs}. The first relation in Eq.~\eqref{eq:vaceqs} determines the form of this function and implies, at leading order, the equality
\begin{equation}\label{eq:secondeq}
\alpha_0r^{p}+n\beta_0 f_{n}r^{n+q-1}=0\,.
\end{equation}
Matching the exponents gives
\begin{equation}\label{eq:ny}
p=n+q-1\,,
\end{equation}
while matching the coefficients results in
\begin{equation}\label{eq:coeff1}
\alpha_0+n\beta_0 f_{n}=0\,.
\end{equation}

The two equations~\eqref{eq:exp2} and ~\eqref{eq:ny} contain the same information and are thus redundant. On the other hand, for Eqs.~\eqref{eq:coeff2} and~\eqref{eq:coeff1} to hold simultaneously, the following relation must be satisfied,
\begin{equation}\label{eq:nqrelvac}
n=-q,
\end{equation}
so that, in combination with Eq.~\eqref{eq:exp2}, it must be
\begin{equation}
p=-1.    
\end{equation}
One immediate consequence is that there is no odd value of $q$ for which $n$ is even and, therefore, such that vacuum spacetimes are regular in the sense of Eq.~\eqref{eq:regtay}. Hence, no theory in the ZK class with a four-dimensional interpretation, and thus satisfying Eq.~\eqref{eq:refalphbet}, can yield geodesically complete vacuum solutions (although curvature invariants may be bounded). This feature has been noted before~\cite{Arrechea:2026ngi}. Solutions of these theories in the presence of matter may be regular, as discussed below.

Beyond interpreting specific metrics as vacuum solutions of specific instances of Eq.~\eqref{eq:sphnew}, these field equations allow us to study how these regular black hole metrics change when matter fields are introduced. This problem has not been studied before in the literature, in particular due to the lack of a set of equations providing a suitable framework.\footnote{The closest framework available was that of nonlinear electrodynamics~\cite{Ayon-Beato:1998hmi,Ayon-Beato:1999kuh,Bronnikov:2000yz,Bronnikov:2000vy}, which is nevertheless different as this type of regularization can only regularize a single vacuum solution with a specific value of the asymptotic mass~\cite{Huang:2025uhv}, while in the construction followed here all vacuum solutions can be regularized.}

\section{Scalar field}\label{sec:scalar}

In this section we discuss two situations when the source $T_{\mu\nu}$ in Eq.~\eqref{eq:sphnew} corresponds to a scalar field, focusing on theories in the ZK family.

The first section below discusses static solutions that generalize known solutions describing naked singularities in general relativity~\cite{Fisher:1948yn}.
The second section deals with time-dependent monopole perturbations of a black hole due to a scalar field.

For simplicity, in most of the equations below we will omit the variables on which the functions depend as these should be contextually clear, for instance writing $\bm{\alpha}$ instead of $\bm{\alpha}[r,f(t,r)]$.

\subsection{Static solutions}\label{sec:stascal}

Inserting in Eq.~\eqref{eq:sphnew} the static and spherically symmetric ansatz in Eq.~\eqref{eq:metric_back} and the stress-energy tensor of a static scalar field $\Phi(r)$ with mass $m$, we obtain the following differential equations:
\begin{equation}
    \frac{ h \left[ \bm{\alpha} + \bm{\beta} f' - 8 \pi r^{2}\left(\,m^{2} \,\Phi^{2} + \Phi'^{2} f\right)\right]}{2 r^{2}}=0,
    \label{eq:background10}
\end{equation}
\begin{equation}
    -\frac{\left(\bm{\alpha} - 8 \pi m^{2} r^{2}\Phi^{2}\right)/f+ \bm{\beta}h'/h  +2(\partial_r\bm{\beta} - \partial_\chi\bm{\alpha})  + 8 \pi r^{2} \Phi'^{2}}{2 r^{2}}=0,
    \label{eq:background20}
\end{equation}
\begin{equation}
\begin{aligned}
    \frac{1}{8}  r\Bigg\{f \bigg[32 \pi r \Phi'^{2}- \bm{\beta}\bigg(\frac{h'^2}{h^2} + \frac{f'h'}{fh}+ 2 \frac{h''}{h}-2 \frac{h'}{h^2}\bigg)
-4 \partial^2_{rr}\bm{\beta}- 2\bigg(\frac{h'}{h}+\frac{f'}{f}\bigg)
\big(2 \partial_r\bm{\beta} - \partial_\chi\bm{\alpha}\big)+&\\+ 4\partial^2_{r\chi}\bm{\alpha}
\bigg]
+2  \left(16 \pi m^{2} r\, \Phi^{2}
- \partial_r\bm{\alpha}\right)-2 f'f\Big(
\frac{h'}{h} \partial_\chi\bm{\beta}
+ 2 \partial^2_{r\chi}\bm{\beta}
- 2 \partial^2_{\chi\chi}\bm{\alpha\Big)}\Biggl\}=0,\label{eq:background30}&
\end{aligned}
\end{equation}
with $\chi=f(r)$. The Klein-Gordon equation for the scalar field reads:
\begin{equation}\label{eq:KGeq}
    -\frac{m^2 \Phi}{f} +\frac{\left(4+ r h'/h+r f'/f\right) \Phi'}{2r}+\Phi''=0.
\end{equation}
In the case in which $\bm{\alpha}$ and $\bm{\beta}$ are those given in Eq.~\eqref{eq:grchoicesalphabeta} we recover general relativity for this particular matter content, which was analyzed in~\cite{Fisher:1948yn} for instance.

Assuming $m=0$, we can solve the Klein-Gordon equation for $\Phi'(r)$, Eq.~\eqref{eq:KGeq}, obtaining
\begin{equation}
    \Phi'(r)=\frac{ c_1\, }{r^2 \sqrt{f(r) h(r)}}.
\end{equation}
Substituting this value into the differential equations above~(\ref{eq:background10}-\ref{eq:background30}), solving the first for $h'(r)$ and the second for $f'(r)$, and then substituting these results into the third equation, shows that the latter is automatically satisfied. The only two independent equations remaining are:
\begin{equation}
    h r\left(\bm{\alpha}+\bm{\beta}f'\right)=\frac{8\pi c_1^2 }{r},
    \label{eq:background1}
\end{equation}
\begin{equation}
      hr\left[\bm{\alpha}+\bm{\beta}\frac{h' f}{h}
       {+2 f ( \partial_r\bm{\beta}- \partial_\chi\bm{\alpha})}
      \right]=-\frac{8\pi c_1^2 }{r}.
      \label{eq:background2}
\end{equation}
These equations further simplify for the theories we will be focusing on, satisfying Eq.~\eqref{eq:intcond} and~\eqref{eq:alphaZK}, which we will assume in the following.  

We can solve these equations perturbatively as for the vacuum case. Considering first Eq.~\eqref{eq:background1}, we obtain:

\begin{equation}\label{ExpnaStatic}
   \alpha_0 r^p+n f_n \beta_0 r^{n+q-1}=\frac{8 \pi c_1^2}{h_0} r^{-2},
\end{equation}
which implies that
\begin{equation}\label{eq:pqnsca}
 p=q+n-1=-2.
\end{equation}
On the other hand, subtracting Eq.~\eqref{eq:background2} from Eq.~\eqref{eq:background1} we obtain that $(fh)'=0$, thus $h(r)=C /f(r)$ and therefore
\begin{equation}\label{eq:pqnsca2}
m=n.  
\end{equation}

A direct consequence of Eqs.~\eqref{eq:pqnsca} and~\eqref{eq:pqnsca2} is:
\begin{equation}\label{eq:nqrelsca}
 n=-q-1.
\end{equation}
Hence, there exist solutions with bounded curvature at $r=0$ (for which $n\geq2$) for odd values of $q$. Indeed, any odd integer number $q\leq -3$ results in solutions with bounded curvature in the presence of a static scalar field distribution.

A plain comparison between Eq.~\eqref{eq:nqrelsca} and the equivalent relation obtained for vacuum solutions, Eq.~\eqref{eq:nqrelvac}, seems to point that  the presence of a static scalar field together with regularity $n\geq 2$ would seem to impose stronger constraints on the parameter $q$ defining the specific theory in the ZK family, namely $q\leq -3$, instead of the equivalent condition for vacuum solutions alone, $q\leq-2$. However, this potential difference disappears when imposing that $q$ must be an odd integer, as required by Eq.~\eqref{eq:refalphbet}.

Taking into account both aspects together, the presence of matter actually allows for all the parity constraints of all relevant the functions involved to be mutually satisfied while having bounded curvature invariants, which is not possible in vacuum.

An specific example is provided by the following theory:
\begin{equation} 
\bm{\beta}=-\frac{2 (r^2+\ell^2)^2}{r^3},   \label{eq:regularTheory}
\end{equation}
with $\bm{\alpha}$ determined by Eq.~\eqref{eq:alphaZK}. This theory has $q=-3$ and thus, according to our general discussion of vacuum solutions above, has vacuum solutions with bounded curvature invariants but that are geodesically incomplete. However, the behavior at $r\rightarrow0$ for static solutions in the presence of a scalar field is compatible with geodesic completeness.

\begin{figure}
    \centering
    \includegraphics[width=1.05\linewidth]{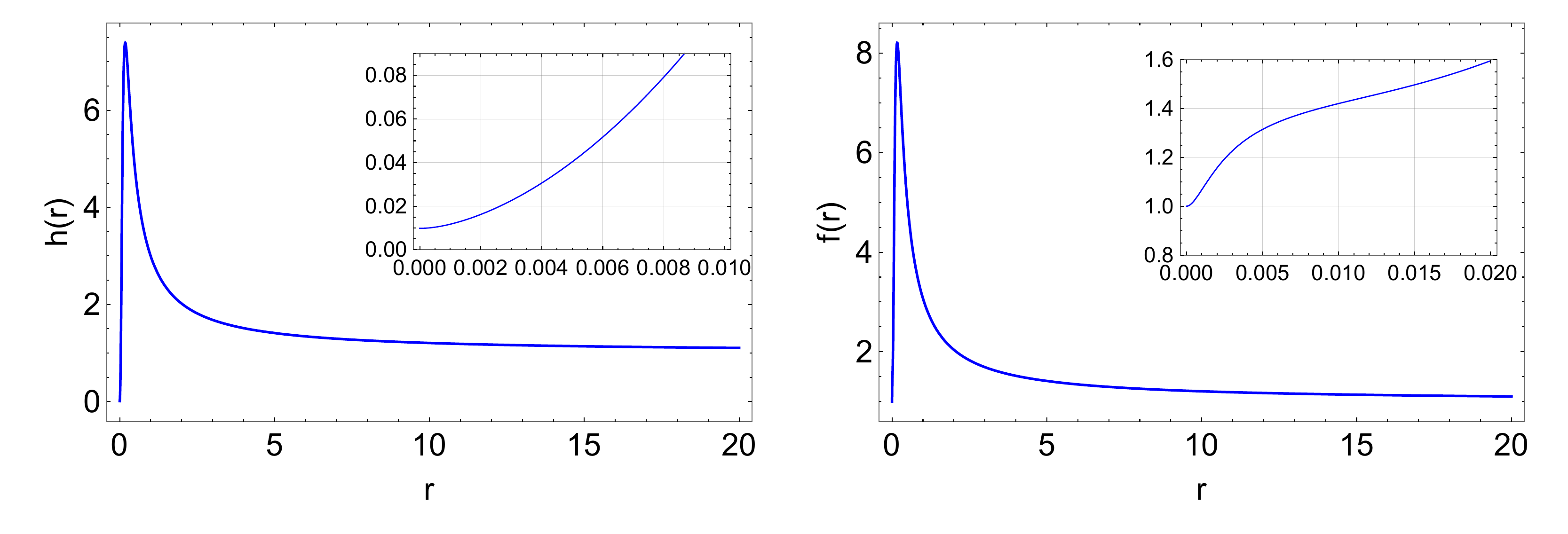}
   
    \caption{Metric components of the solution obtained numerically for $\alpha(\chi,r)=(1-\chi) \partial_r \beta(r)$ and $\beta(r)=-2 (r^2+\ell^2)^2/r^3$ choosing $\ell=10^{-1}$. This solution has bounded curvature invariants for all values of $r$.}
    \label{fig:scalrNum}
\end{figure}

Note that the analytical conditions we derived for $\bm{\alpha}$ and $\bm{\beta}$ are necessary but not sufficient to ensure global regularity, as they guarantee the boundedness of curvature invariants only at $r=0$. To assess the global behavior, we solved Eqs.~\eqref{eq:background1}-\eqref{eq:background2} numerically for this  example, finding indeed a solution without any curvature singularities and without any horizon. The numerical solution is shown in figure \ref{fig:scalrNum}. %Nevertheless, these conditions are more restrictive than those obtained in vacuum, indicating that the presence of matter further constrains the class of theories admitting regular static solutions.

\subsection{Monopole time-dependent gravitational perturbations sourced by a scalar field }
In this section, we delve into the study of linear gravitational perturbations of regular black hole spacetimes sourced by the presence of a time-dependent scalar field perturbation. Our specific aim here is to identify the conditions required to maintain spacetime regularity dynamically even when matter-induced perturbations are introduced. Related previous research includes~\cite{Coviello:2025pla,Spina:2024npx,Daghigh:2020mog,Meng:2022oxg,Franzin:2023slm}.

Due to the spherical symmetric ansatz used to derive the equations of motion~\eqref{eq:sphnew}-\eqref{eq:geneqs2}, we are restricted to study only the polar monopole modes (e.g.,~\cite{Berndtson:2007gsc,Brito:2016peb}), where the perturbations $\delta q_{ab}(x)$ and $\delta r(x)$ can be parameterized as:
\begin{equation}
\delta g_{\mu\nu}\text{d}y^\mu\text{d}y^\nu=f(r)H_0(t,r)\text{d}t^2+2H_1(t,r)\text{d}t\text{d}r+\frac{H_2(t,r)}{f(r)}\text{d}r^2+r^2K(t,r)\gamma_{ij}\text{d}\theta^i\text{d}\theta^j.
\end{equation}
For such a perturbed metric, $\chi$ reads:
\begin{equation}
\chi= f + \epsilon e^{-i \omega t} f (-H_2 + K_0 + r K_0') .
\end{equation}
Previous works using a similar parameterization but in other frameworks and having a different scope can be found in~\cite{Bronnikov:1979uz,Martel:2005ir,Bronnikov:2012ch,Bronnikov:2023amc,Mukohyama:2025owu}.

We will evaluate gravitational perturbations up to linear order in $\epsilon$, sourced by scalar fluctuations of amplitude $\sqrt{\epsilon} $, as the scalar stress energy tensor is quadratic in the field. Consequently, the leading order contribution to the master field equations in Eq.~\eqref{eq:sphnew} is of order $\epsilon$, while the Klein-Gordon equation is dominated by terms of order $\sqrt{\epsilon}$. We decompose the scalar field as
\begin{equation}
\Phi(r,t)=\Phi(r)\exp(-i \omega t/2)    
\end{equation}
and the gravitational perturbations as
\begin{equation}
H_i(t,r)= H_i(r) \exp(- i \omega t),\quad K(t,r)=K(r)\exp(-i\omega t).
\end{equation}

Up to order $\sqrt{\epsilon}$, the Klein-Gordon equation reads
\begin{equation}
\label{eq:KG}
 \sqrt{\epsilon} \left[\left(- \frac{\bm{\alpha}}{\bm{\beta}}+\frac{2 f}{r}\right) \Phi'+ f  \Phi ''+\frac{\omega ^2
   \Phi  }{4 f}\right]=0,
\end{equation}
corresponding to the equation of a test scalar field evolving on the regular background. The quasinormal modes for this system have been computed (e.g., in \cite{Flachi:2012nv,Franzin:2023slm}) and no growing modes have been found. 

Focusing in the gravitational sector, we impose the Lorenz gauge: 
\begin{equation}
    \nabla^\beta \left( h_{\rm \alpha\beta}-\frac{1}{2} g_{\alpha \beta} h\right)=0.
    \label{eq:Lorenz-gauge}
\end{equation}
With this gauge choice, $H_2$ and $H_0$ can be expressed in terms of $H_1$ and $K_0$, which remain our only two independent degrees of freedom. The gauge conditions in Eq.~\eqref{eq:Lorenz-gauge} indeed impose: 

\begin{align}
H_0 = -&\frac{i}{2 \omega \bm{\beta}^3 r f } \Bigg\{
\bm{\beta}^3 \Big[ \omega^2 H_1 r^2 + 2 i \omega r f  (3 K_0 + r K_0') + f^2 (6 H_1 + 6 r H_1' + r^2 H_1'') \Big]- \bm{\alpha}^2 f H_1 r^2 \partial_\chi \bm{\beta} \notag \\
&- \bm{\beta}^2 r \Big[ \bm{\alpha} (8 f H_1 + i \omega K_0 r + 3 r f  H_1') + f H_1 r \partial_r \bm{\alpha} \Big] +\bm{\alpha} \bm{\beta} H_1 r^2 \Big( \bm{\alpha} + 2 f \partial_\chi \bm{\alpha} \Big)  \Bigg\}
\end{align}
and
\begin{align}
H_2 = &\frac{i}{2 \omega \bm{\beta}^3 r f } \Bigg\{ \bm{\beta}^3 \Big[ -\omega^2 H_1 r^2 - 2 i \omega r f  (K_0 + r K_0') + f^2 (2 H_1 - 2r H_1' - r^2 H_1'') \Big] + \bm{\alpha}^2 f H_1 r^2 \partial_\chi \bm{\beta}\notag\\
&+ \bm{\beta}^2 r \Big[ \bm{\alpha} (4 f H_1 + i \omega K_0 r + 3 r f  H_1') + f H_1 r \partial_r \bm{\alpha} \Big]- \bm{\alpha} \bm{\beta} H_1 r^2 \Big( \bm{\alpha} + 2 f \partial_\chi \bm{\alpha} \Big)  \Bigg\}.
\end{align}

Only two of the four field equations in Eq.~(\ref{eq:sphnew}-\ref{eq:geneqs2}) are independent, and we will work with the independent equations $\mu=1$, $\nu=2$ and $\mu=\nu=2$. In the Lorenz gauge, these equations are expressed in terms of the only two independent functions and read, up to order $\epsilon$:

\begin{align}
\label{eq:pert1}
\frac{\epsilon e^{-i \omega t} }{4 \bm{\beta}^2 r^3 f } \Bigg\{&\bm{\beta}^3 \Big[ \omega^2 H_1 r^2 + f^2 (-2 H_1 + 2 r H_1' + r^2 H_1'') \Big] + \bm{\alpha} \bm{\beta} H_1 r^2 (\bm{\alpha} + 2 f \partial_\chi \bm{\alpha})- \bm{\alpha}^2 f H_1 r^2 \partial_\chi \bm{\beta}\notag \\
&- \bm{\beta}^2 r \Big[ \bm{\alpha} (2 i \omega K_0 r + 4f H_1 + 3f r H_1') + f r (-16 i \pi \omega r \Phi \Phi' + H_1 \partial_r \bm{\alpha}) \Big] \Bigg\}=0
\end{align}

and

\begin{align}
\label{eq:pert2}
\frac{ \epsilon e^{-i \omega t}}{2 \omega \bm{\beta}^2 r^3 f^2 } \Bigg\{\omega \bm{\beta}^3 r \Big[\omega^2 K_0 r + f^2 (4 K_0' + r K_0'')\Big]&
+\omega \bm{\beta} r \Big[ \bm{\beta}r  f \big(  -32 \pi f \Phi \Phi' + K_0 \partial_r \bm{\alpha} \big) + 16 \pi r^2 f^2  \Phi \Phi' \partial_\chi \bm{\alpha} \Big]
\notag \\
+ \bm{\alpha} f \Big[ \bm{\beta} \big(
- 2 \omega r^2 (8 \pi r \Phi \Phi' + K_0 \partial_\chi \bm{\alpha}) \big) - 16 \pi & \omega r^3 f  \Phi \Phi' \partial_\chi \bm{\beta} \Big] 
+ \bm{\alpha}^2 r \Big[ \bm{\beta} (2 i f H_1 + \omega K_0 r) + \omega f K_0 r \partial_\chi \bm{\beta} \Big]
\notag\\
+ \bm{\alpha}\bm{\beta}^2 f \Big[-2 i f (&2 H_1 + r H_1') + \omega r (5 K_0 + r K_0')  \Big]
\Bigg\}=0.
\end{align}

In deriving these equations, we assumed that the background equations~\eqref{eq:vaceqs} are satisfied. Furthermore, since we set $h(r)=f(r)$, these equations imply the validity of Eq. \eqref{eq:intcond}. Now, we focus on theories in the ZK class satisfying Eq.~\eqref{eq:alphaZK}, which in particular implies $\partial_\chi\bm{\beta}=0$.
We solved such equations, together with the Klein Gordon equation, perturbatively near 0, assuming the following expansion for the perturbations functions and scalar field:
\begin{equation}\label{eq:pertexps}
    H_1(r)=\sum_{i=0}^\infty \mathcal{H}_ir^i,\quad K_0(r)=\sum_{i=0}^\infty \mathcal{K}_ir^i,\quad \Phi(r)=\sum_{i=s}^\infty \varphi_ir^i.
\end{equation}
The expansions of $H_1$ and $K_0$ start at $i=0$ since we require the metric perturbations to be regular at $r=0$: if they diverged there, they would not constitute a valid perturbation of the background and the perturbative expansion would break down. For the scalar field, which has no background value, the leading power $s$ is instead determined by the Klein Gordon equation.

After a lengthy calculation, the steps of which are summarized in Appendix~\ref{app:perturbative}, we found that requiring regularity of the Kretschmann scalar at order $\epsilon$ imposes the condition
\begin{equation}
    q\leq -3.
\end{equation}

Hence, we find the same conditions for time-dependent scalar perturbations of a black holes as for a static configuration supported by a scalar field. The same considerations discussed in the static scalar case thus apply to the time-dependent situation. This is an indication that the presence of dynamically evolving matter fields does not lead to additional constraints as compared to static matter fields. Hence in the following two sections, in which we study matter fields with a more complex structure, we will restrict our attention to static configurations for simplicity.

\section{Electromagnetic field}\label{sec:maxwell}

We now focus on electrovacuum solutions in the ZK class (although some of the expressions below are more general~\cite{Carballo-Rubio:2026mvj}). Specifically, we seek solutions to a generalization of the Einstein--Maxwell equations where the gravitational field equations take the form in Eq.~\eqref{eq:sphnew}. The associated stress-energy tensor is given by
\begin{equation}
T_{\mu\nu}=\frac{1}{4\pi}\left(F_{\mu\alpha}F_\nu^{\ \alpha}-\frac{1}{4}g_{\mu\nu}F_{\alpha\beta}F^{\alpha\beta}\right),    
\end{equation}
while the electromagnetic field equations are
\begin{equation}\label{eq:maxeqs}
\nabla_\nu F^{\nu\mu}=\frac{1}{\sqrt{-g}}\partial_\nu\left(\sqrt{-g}F^{\nu\mu}\right)=4\pi J^\mu,    
\end{equation}
with the source term
\begin{equation}\label{eq:maxeqssource}
J^t=Q\delta(r).    
\end{equation}
For the metric ansatz from Eq.~\eqref{eq:metric_back},
 the field equations reduce to the following equations (see~\cite{Carballo-Rubio:2026mvj} for a step-by-step discussion):
\begin{align}
\frac{h(\bm{\alpha}+\bm{\beta}f')}{2r^2}&=\frac{fQ^2}{r^4},\nonumber\\
\frac{\bm{\alpha}/f+\bm{\beta}h'/h-2\left(\partial_\chi\bm{\alpha}-\partial_r\bm{\beta} \right)}{2r^2}&=\frac{Q^2}{h r^4}. 
\end{align}
We once more focus on theories satisfying the integrability condition in Eq.~\eqref{eq:intcond}. In this case, the two equations become dependent, leaving us with a single equation:
\begin{equation}\label{eq:maineqmax}
\bm{\alpha}+\bm{\beta}f'=\frac{2 Q^2}{r^2}.    
\end{equation}
While this equation can be solved exactly, as discussed in~\cite{Carballo-Rubio:2026mvj}, for the sake of analyzing the ZK family and comparing with the previous scalar field case, we perform a perturbative expansion around $r=0$ as described in Sec.~\ref{sec:zkexp}. Substituting the expansions in Eqs.~\eqref{eq:fexp},~\eqref{eq:hexp} and ~\eqref{eq:taylor}, we have:
\begin{equation}\label{ExpnaMaxwell}
   \alpha_0 r^p+n f_n \beta_0 r^{n+q-1}=2Q^2 r^{-2}.
\end{equation}
It is interesting to note that this equation has a similar form as Eq.~\eqref{ExpnaStatic} valid for a static scalar field. It thus implies the same relation
\begin{equation}\label{eq:pqnscaMaxwell}
 p=q+n-1=-2.
\end{equation}

Consequently, solutions with a regular Kretschmann scalar, which require $n\geq 2$, are admitted under exactly the same conditions as in the case of a static scalar field:
\begin{equation}
    q\leq -3, \qquad p=-2.
\end{equation}
Note that, as observed in the static scalar field and contrary to the vacuum case, theories with odd values of $q$ lead to even values of $n=-q-1$.

As an example of a charged regular black hole, we consider the same theory that was shown to yield a regular static solution in the presence of a scalar field, which $\bm{\beta}$ given by Eq.~\eqref{eq:regularTheory} and $\bm{\alpha}$ determined from the latter by Eq.~\eqref{eq:alphaZK}. Introducing an electromagnetic field, Eq.~\eqref{eq:maineqmax} can be solved, resulting in the following metric:
\begin{equation}
      f(r)=1- \frac{ r^2( 2 M r- Q^2)}{(r^2 + \ell^2)^2}.
\end{equation}
For $\ell=0$, this metric reduces to the classical Reissner-Nordstr\"om metric, while for $\ell\neq0$ describes a black hole spacetime with bounded curvature which, however, is not geodesically complete due to $f(r)$ not being an even function on $r$~\cite{Zhou:2022yio,Arrechea:2026ngi}. A more detailed discussion of the landscape of theories and solutions for the charged case is provided in~\cite{Carballo-Rubio:2026mvj} (see also the related discussion in~\cite{Colleaux:2026hat}).

For completeness, it is interesting to note that the geometry for a charged regular black hole constructed \emph{ad hoc} by Frolov in~\cite{Frolov:2016pav} does not arise as a solution of the field equations described in this section~\cite{Carballo-Rubio:2026mvj}, and thus its interpretation as a genuinely charged solution is unclear. We also note that, while in this paper we are only considering modifications in the dynamics of the gravitational sector and not of the Maxwell sector, studying modifications in the dynamics of the latter can also lead to interesting features~\cite{2026arXiv260310110B,Cano:2020ezi,Cano:2020qhy,Hennigar:2025yqm,PinedoSoto:2026hfm,2026arXiv260310110B}.

%%%%%%%%%%%%%%%%%%%%
\section{Series expansion of static solutions with general matter contents}\label{sec:gensta}
%%%%%%%%%%%%%%%%%%%%

To close the discussion of the ZK family, in this section we make explicit that the static solutions above (for a scalar and a electromagnetic field, discussed in Secs.~\ref{sec:stascal} and~\ref{sec:maxwell}, respectively) are particular cases of situations in which the static matter source can have more general behaviors in the $r\rightarrow0$ limit.

For a general static matter content, and theories satisfying the integrability condition in Eq.~\eqref{eq:intcond}, the field equations to be solved are:
\begin{align}
\frac{h\left(\bm{\alpha}+\bm{\beta}f'\right)}{2r^2}&=8\pi T_{tt},\nonumber\\
-\frac{\bm{\alpha}h+\bm{\beta}fh'}{2r^2 fh}&=8\pi T_{rr}.
\end{align}
These equations generally yield two different functions $h(r)$ and $f(r)$. However, for stress-energy tensors satisfying the following relation,
\begin{equation}\label{eq:resetrel}
T_{rr}=-\frac{1}{fh}T_{tt},    
\end{equation}
the metric functions are constrained to satisfy
\begin{equation}\label{eq:singfunc}
hf'=fh',    
\end{equation}
and thus 
both functions are the same up to an arbitrary integration constant that can be absorbed in a redefinition of the time coordinate. Both the scalar field and electromagnetic field cases satisfy Eq.~\eqref{eq:resetrel} in static situations. 

For such situations in which Eqs.~\eqref{eq:resetrel} and~\eqref{eq:singfunc} are satisfied, we are left with a single equation
\begin{equation}
\bm{\alpha}+\bm{\beta}\partial_r f=S,
\end{equation}
where, for simplicity, we have defined the source term as
\begin{equation}\label{eq:Tdef}
S=\frac{16\pi r^2}{f}T_{tt}.   
\end{equation}
The behavior of this quantity (that is proportional to the energy density) in the $r\rightarrow0$ limit depends on the specific matter content considered, as determined by the field equations satisfied by the matter fields. We can generically parametrize the source as:
\begin{equation}
S=r^w\sum_{u=0}^\infty s_u r^u.
\end{equation}
Instead of Eq.~\eqref{eq:secondeq} obtained in vacuum situations, we thus obtain
\begin{equation}
r^p\alpha_0+n\beta_0f_n r^{n+q-1}=r^ws_0.
\end{equation}
As a consequence, we have the relations
\begin{equation}
p=n+q-1=w,
\end{equation}
and
\begin{equation}
\alpha_0+n\beta_0f_n=s_0.
\end{equation}

These relations reduce to Eqs.~\eqref{eq:ny} and~\eqref{eq:coeff1} whenever the source vanishes. However, the limit is not continuous as, whenever there is a source, it follows that $p$ is specified by its leading-order behavior,
\begin{equation}
p=w,    
\end{equation}
while the leading-order behavior of the metric is constrained to satisfy
\begin{equation}\label{eq:regcoeffsmat}
n=-q+w+1,    
\end{equation}
which is to be compared with $p=-1$, $n=-q$ obtained in the vacuum (no source) case.

Hence, bounded curvature requires
\begin{equation}
q\leq w-1.
\end{equation}
In general relativity, $q=1$ and, assuming that the stress-energy tensor is quadratic in the fields and taking into account the $r^2$ factor in Eq.~\eqref{eq:Tdef}, we generally have $w=2$ and thus the inequality above is satisfied. However, any singular behavior of these fields that results in $w\leq 1$ cannot lead to regular geometries in general relativity. However, lower values of $q$ can regularize the gravitational fields of such singular matter fields, as we have seen in the examples above characterize by $w=-2$, which requires $q\leq -2$.

It is also interesting to note that, for a fixed value of $w$, a lower value of $q$ implies a higher value of $n$ according to Eq.~\eqref{eq:regcoeffsmat}. For instance, regular sources with $w=2$ that result in $n=2$ for $q=1$ (general relativity), have instead $n=6$ for $q=-3$. Hence, the starting powers in the expansions of the metric functions around their constant values at $r\rightarrow0$, recall Eqs.~\eqref{eq:fexp} and~\eqref{eq:hexp}, are determined by the degree of regularization of the dynamics of the gravitational field. This difference in the behavior of metric functions around the center of symmetry is likely to be an important feature in analyses of dynamical solutions.

\section{Homogeneus and isotropic massless perfect fluid}\label{sec:cosm}

We now focus on theories in the KMT class. As discussed in Sec.~\ref{sec:spaces}, for these theories we will be working with a different matter content that allows us to work with the field equations without resorting to a perturbative expansion around $r=0$ as in previous sections.

In general relativity, the homogeneous and isotropic solution of the field equations in vacuum is regular and corresponds to Minkowski spacetime. This solution becomes singular when adding matter, for instance in the presence of a massless perfect fluid corresponding to the radiation present in the early universe. 

In this section, we study the homogeneous and isotropic solutions for theories in the KMT class that also have Minkowski spacetime as a solution in vacuum. Related previous work include~\cite{Chinaglia:2017wim,Bueno:2025gjg,Ling:2025ncw,Borissova:2026klg}.

We consider the following metric ansatz
\begin{equation}
g_{\mu\nu}(y)\text{d}y^\mu\text{d}y^\nu=-\text{d}t^2+\frac{a(t)^2}{1-KR^2}\text{d}R^2+a(t)^2 R^2\text{d}\Omega^2,
\end{equation}
where we are using coordinates $\{x^0,x^1\}=\{t,R\}$. The connection with the notation introduced above can be made explicit through the identifications
\begin{equation}
q_{ab}(x)\text{d}x^a\text{d}x^b=-\text{d}t^2+\frac{a(t)^2}{1-KR^2}\text{d}R^2,\quad r(x)=a(t)R,
\end{equation}
while 
\begin{equation}
\chi=1-R^2\left(K+\dot{a}^2\right).    
\end{equation}
The general field equations take the form:
\begin{align}
\mathscr{G}_{tt}&=\frac{\bm{\alpha}a-2\bm{\beta}R\left(K+\dot{a}^2\right)}{2R^2a^3}+\dot{a}^2\frac{\partial_\chi\bm{\alpha}-\partial_r\bm{\beta}}{a^2} =8\pi\rho,\nonumber\\
\mathscr{G}_{tR}&=\dot{a}aR\left(\partial_\chi\bm{\alpha}-\partial_r\bm{\beta} \right)=0, \nonumber\\
\mathscr{G}_{RR}&=-\frac{\bm{\alpha}-2\bm{\beta}R\ddot{a}}{2R^2\left(1-KR^2\right)}+\frac{\partial_\chi\bm{\alpha}-\partial_r\bm{\beta}}{R^2}=\frac{a^2}{1-KR^2}8\pi p.
\end{align}
The standard equations for cosmological spacetimes in general relativity are recovered for the choices in Eq.~\eqref{eq:grchoicesalphabeta}, namely $\bm{\alpha}^{({\rm GR})}=2\left(1-\chi\right)=2R^2\left(K+\dot{a}^2\right)$ and $\bm{\beta}^{({\rm GR})}=-2r=-2Ra$.

To determine the form of $\bm{\alpha}$ and $\bm{\beta}$ compatible with homogeneity and isotropy, we require $\mathscr{G}_{\mu\nu}$ to be invariant under the spatial Killing vectors of the FLRW geometry. Rotational invariance is already guaranteed by spherical symmetry, while invariance under the translational Killing vectors\footnote{Note that for translations the Killing vectors are $
    \xi_{1}=\sqrt{1-KR^2}\left(\sin\theta\cos\phi\partial_R+\frac{\cos\theta\cos\phi}{R}\partial_\theta-\frac{\sin\phi}{R\sin\theta}\partial_\phi\right)$, $\xi_{2}=\sqrt{1-KR^2}\left(\sin\theta\sin\phi\partial_R+\frac{\cos\theta\sin\phi}{R}\partial_\theta+\frac{\cos\phi}{R\sin\theta}\partial_\phi\right)$ and $\xi_3=\sqrt{1-KR^2}\left(\cos\theta\partial_R-\frac{\sin\theta}{R}\partial_\theta\right)$. Those of rotations are instead $\zeta_1=\sin\phi \partial_\theta+\cot\theta\cos\phi\partial_\phi$, $\zeta_2=-\cos\phi \partial_\theta+\cot\theta\sin\phi\partial_\phi$ and $\zeta_3=\partial_\phi$.
    }, after factoring out the fixed coordinate dependence of the FLRW metric, imposes the absence of any additional dependence on the radial coordinate $R$.
From the definitions of $r$ and $\chi$, we note that the combination
\begin{equation}\label{eq:psidef}
\psi=\frac{1-\chi}{r^2}=\frac{K+\dot a^2}{a^2}
\end{equation}
is independent of $R$.  Assuming non static universes (the $tR$ equation imposes the integrability condition) and imposing the Killing invariance on the diagonal components of $\mathscr{G}_{\mu\nu}$ requires the combinations $\bm\alpha/r^2$ and $\bm \beta/r$ to depend on $r$ and $\chi$ only through this spatially homogeneous variable: 
\begin{equation}
    \frac{\bm\alpha}{r^2}=\bm{N}(\psi), \qquad \frac{\bm \beta}{r}=\bm{M}(\psi).
\end{equation}
Requiring Minkowski spacetime (that has $\psi=0$) to be a regular vacuum solution implies $\bm{N}(0)=0$. We can therefore write $\bm{N}(\psi)=2 \psi \bm A(\psi)$, while defining $\bm{M}(\psi)=-2\bm B(\psi)$, where the numerical factors have been chosen so that general relativity corresponds simply to $\bm A=\bm B=1$. Hence,
\begin{equation}\label{eq:AalphaBbeta}
\bm{\alpha}=2(1-\chi)\bm{A}\left(\frac{1-\chi}{r^2}\right),\quad \bm{\beta}=-2r\bm{B}\left(\frac{1-\chi}{r^2}\right).
\end{equation}
For completeness, we note that these relations cannot be satisfied by the theories characterized by the ZK family, characterized by $\bm{\Sigma}=1$ in Eq.~\eqref{eq:zkmtdef}.\footnote{Note that this does not necessarily imply that these theories are not observationally viable, as it only imposes a minimum degree of inhomogeneity and isotropy given by the new length scale in these theories, which can be arbitrarily small, although a dedicated analysis of this issue is lacking in the literature.} However, these relations are compatible with the KMT family, characterized by $\bm{\Phi}=1$ in Eq.~\eqref{eq:zkmtdef}, which are the focus in this section. 

On the other hand, in consonance with the rest of our analysis we impose the integrability condition in Eq.~\eqref{eq:intcond}, which implies the additional relation for the functions $\bm{A}$ and $\bm{B}$:
\begin{equation}
\bm{A}+\psi\bm{A}'=\bm{B}-2\psi\bm{B}',
\label{eq:constraint_AB}
\end{equation}
where the derivatives are with respect to the new variable $\psi$ defined in Eq.~\eqref{eq:psidef}.

In terms of $\bm A$ and $\bm B$, the Friedmann equation is written as
\begin{equation}\label{eq:friedmanneq}
\frac{\left(1-\chi\right)\left(\bm{A}+2\bm{B}\right)}{R^2a^2}=\left( H^2+\frac{K}{a^2}\right)\left(\bm{A}+2\bm{B}\right)=8\pi\rho,    
\end{equation}
where we are using the standard definition $H\equiv \dot{a}/a$. Eq.~\eqref{eq:friedmanneq} suggests introducing the definition
\begin{equation}\label{eq:G_def}
    \bm{C}(\psi) \equiv \bm{A}(\psi) + 2\bm{B}(\psi)\,,
\end{equation}
so that the modified Friedmann equation takes the form 
\begin{equation}\label{eq:mod_friedmann}
 \psi \, \bm{C}(\psi) = 8\pi\rho.
\end{equation}
%
%In general relativity, both $\bm{A}$ and $\bm{B}$ reduce to the unit constant.

On the other hand, the constraint in Eq.~\eqref{eq:constraint_AB} is identically satisfied by the expressions:
\begin{equation}\label{eq:A_B_from_G}
    \bm{A} = \frac{\bm{C} - 2\psi\,\bm{C}'}{3}\,, \qquad 
    \bm{B} = \frac{\bm{C} + \psi\,\bm{C}'}{3}\,.
\end{equation} 

\subsection{Homogeneous and isotropic vacuum solutions}
As we said, the homogeneous and isotropic solution in vacuum is regular already in general relativity. For other theories in the KMT class, the conditions for regularity are mild. It is reasonable to ask that the vacuum solution is in this case the flat spacetime (Minkowski: $a=a_0$, $\dot{a}=0$). The modified Friedmann equation in this case reads:
\begin{equation}
   \psi\, \bm{C}(\psi) =0.
\end{equation}
For this equation to have $\dot{a}=0 $ ($\psi=H^2=0$) as a solution, the following behavior of $\bm C$ is required:
\begin{equation}\label{eq:ccondsmin}
    \bm{C}(\psi) \xrightarrow{\psi\rightarrow0} \psi^v,  \qquad v>-1.
\end{equation}
Through equations \eqref{eq:A_B_from_G} and \eqref{eq:AalphaBbeta} this translates into:
\begin{equation}
   \bm{ \alpha} \xrightarrow{\psi\rightarrow0} \psi^{v+1}, \qquad{} \bm{\beta}\xrightarrow{\psi\rightarrow0} \psi^v, \qquad v>-1. 
\end{equation}

\subsection{Massless perfect fluid solutions}

In the following, we will consider a massless perfect fluid source satisfying
\begin{equation}\label{eq:radiation_eos}
    p = \frac{\rho}{3}\,, \qquad \rho = \frac{\rho_0}{a(t)^4}\,.
\end{equation}
\subsubsection{Bouncing cosmologies}
 
A smooth cosmological bounce characterized by $\dot{a} = 0$ (i.e., $H = 0$) for some finite $a = a_\mathrm{min} > 0$ requires that $ \lim_{H \to 0} H^2\, \bm{C}(H^2) =8\pi\rho_0/a_\mathrm{min}^4 > 0$, where we have used explicitly that $\lim_{K\rightarrow0}\psi=H^2$. This is incompatible with the condition that Minkowski spacetime (for which $\dot{a}=0$ identically) must be a vacuum solution, which implies $ \lim_{H \to 0} H^2\, \bm{C}(H^2) =0$.

We can therefore conclude that, for $K = 0$ (flat universe), within the class of theories  that admit Minkowski spacetime as a vacuum solution and have field equations of up to second order in the derivatives of the metric, there exist no bouncing cosmological solutions sourced by radiation (or any fluid with $\rho > 0$). Note that we are assuming  $\bm{C}(\psi)$ to be a single-valued function, which is connected to the existence of a single action. Considering that $\bm{C}(\psi)$ is multi-valued leads to multiple branches of solutions and can evade this argument if requiring that Minkowski spacetime is not a solution for at least one of the branches, although this introduces further complications in relation to the interpretation of the resulting non-conventional ``dual-action" structure~\cite{Ling:2025ncw,Borissova:2026klg}.

For $K=1$ instead, bouncing cosmologies can exist~\cite{Bueno:2025gjg}. This is possible because at the bounce $\psi_{\rm bounce}=1/a_{\rm min}^2\neq 0$, and therefore the condition for a bounce becomes $\lim_{\psi\rightarrow 1/a^{2}_{\rm min}}\psi\,\bm{C}(\psi)=8\pi\rho_0/a^4_{\rm min}$, which is not incompatible with the condition required for Minkowski spacetime to be a solution, thus evading the previous no-go argument.

\subsubsection{Regular cosmologies via emergent de Sitter}
 
Although bounces are excluded for $K=0$, regularity can still be achieved if $a \to 0$ only asymptotically as $t \to -\infty$, while the Hubble parameter remains bounded (thus with $\dot{a}\propto a$ in this limit~\cite{Carballo-Rubio:2024rlr,Garcia-Saenz:2024ogr}). This requires $\bm{C}(\psi)$ to diverge at some critical value $\psi_c > 0$, which in turn caps $H$ at $H_\mathrm{max} = \sqrt{\psi_c}$. On the other hand, the condition $\bm{C}(0) = 3$ is required to recover general relativity at low curvatures.

Regarding the divergence of $\bm C$, we need $\bm{C}(\psi) \to \infty$ as $\psi \to \psi_c = 1/\ell^2$ for some length scale $\ell > 0$. In the same limit, we have $r\propto a$ and $\left(1-\chi\right)\propto \dot{a}^2 \propto a^2$. 
From the modified Friedmann equation it follows that, to have regular solutions, $\bm{C}$ must diverge as $\left(\dot{a}a\right)^{-2}\sim a^{-4}$ for $\psi \rightarrow\psi_c$. This is valid for example if $\bm{\alpha}=2 (1-\chi) \bm{A} \sim a^{-2}$ when $a\rightarrow0$, or $\bm{\beta}=-2 r \bm{B}\sim a^{-3}$ in the same limit. However, there is also the case in which both $\bm\alpha$ and $\bm\beta$ diverge because both $\bm A$ and $\bm B$ diverge at least as $a^{-4}$ (and additional terms cancel in $\bm C$). The integrability constraint~\eqref{eq:constraint_AB} allows us to strengthen the preceding conclusions. From Eq.~\eqref{eq:A_B_from_G} and the modified Friedmann equation, we have
\begin{equation}
\mathbf{B}
=\frac{1}{3}
\left[\psi\bm{C}(\psi)\right]'=\frac{8\pi}{3}\rho'.
\end{equation}
For $K=0$ we get $\dot\psi=2H\dot H$, while for the massless perfect fluid introduced in Eq.~\eqref{eq:radiation_eos} $\dot\rho=-4H\rho$.
It follows that
\begin{equation}
\rho'
=\frac{\dot\rho}{\dot\psi}
=-\frac{2\rho}{\dot H},
\end{equation}
and hence
\begin{equation}
\mathbf{B}
=-\frac{16\pi}{3}\frac{\rho}{\dot H}.
\end{equation}
In the asymptotic de Sitter regime,
\begin{equation}
\bm{\beta}
=2r\mathbf{B}
\propto\frac{a^{-3}}{\dot H}.
\end{equation}
Thus, the divergence of $\bm{\beta}$ is necessarily present and is stronger than $a^{-3}$ (since $\dot H\to 0$ as $t\to-\infty$).
The integrability constraint also fixes the leading behavior of $\mathbf{A}$. Indeed, using $\bm{C}=8\pi\rho/\psi$,
we find
\begin{equation}
\frac{\mathbf{B}}{\bm{C}}
=-\frac{2\psi}{3\dot H}\to \infty,
\end{equation}
since $\psi\to\psi_c>0$ and $\dot H\to0$. The relation
$\mathbf{A}=\bm{C}-2\mathbf{B}$
therefore implies
\begin{equation}
\frac{\mathbf{A}}{\mathbf{B}}
=\frac{\bm{C}}{\mathbf{B}}-2
\longrightarrow-2,
\end{equation}
so that $\mathbf{A}\simeq-2\mathbf{B}$.
Hence, $\mathbf{A}$ necessarily diverges as well. Since, for $K=0$, $(1-\chi)\propto a^2$
in the asymptotic de Sitter limit, we obtain
\begin{equation}
\bm{\alpha}
=2(1-\chi)\mathbf{A}
\propto\frac{a^{-2}}{\dot H}.
\end{equation}
Therefore, once the integrability constraint is imposed, both
$\bm{\alpha}$ and $\bm{\beta}$ necessarily diverge. Note that this divergence is required for $\psi\rightarrow\psi_c\neq0$ and thus is compatible with the vacuum condition \eqref{eq:ccondsmin} for  $\psi\rightarrow0$.

Under these conditions, Eq.~\eqref{eq:mod_friedmann} can be inverted to give $H^2 = \mathcal{F}(8\pi\rho_0/a^4)$ for some function $\mathcal{F}$ satisfying $\mathcal{F}(x) \to x/3$ for $x \to 0$ (GR regime) and $\mathcal{F}(x) \to 1/\ell^2$ for $x \to \infty$ (de Sitter regime). The two asymptotic behaviors of the scale factor are:
\begin{equation}\label{eq:asymptotics}
    a(t) \sim \begin{cases}
        a_0 \, e^{t/\ell} & \text{for } t \to -\infty, \\[6pt]
       a_{\infty} \  \!\! t^{1/2} & \text{for } t \to +\infty,
    \end{cases}
\end{equation}
with $a_0$ a real integration constant and $a_\infty^4=32\pi\rho_0/3$.

It is interesting to note that, if we evaluate the equivalent of $p$ and $q$ in this cosmological example for the KMT family, their values would be different than the ones obtained for other matter sources in the ZK family. This is to be expected as it occurs even at the level of vacuum solutions, for instance comparing the values of these coefficients between theories leading to a Bardeen-like solution and a Hayward solution. Hence, while there may be a way to meaningfully compare these different coefficients across the two different families, the comparison is not straightforward.

We finish the section by providing explicit choices of $\bm{C}(\psi)$ that yield regular cosmologies such as, for instance:
\begin{equation}\label{eq:power_law_G}
    \bm{C}_{(u)}(\psi) = \frac{3}{\left(1 - \psi\,\ell^2\right)^u}\,, \qquad u > 0\,,
\end{equation}
which satisfies $\bm{C}_{(u)}(0) = 3$ and has a pole at $\psi_c = 1/\ell^2$. 

The functions in Eq.~\eqref{eq:power_law_G} satisfy the required conditions to have Minkowski solutions in vacuum, Eq.~\eqref{eq:ccondsmin}, for any value of $\ell$, while also satisfying the conditions for regular cosmologies in presence of matter if $\ell\neq0$, as we show in the following. These examples therefore illustrate how having regular solutions in the presence of matter (massless perfect fluids) require additional conditions on the structure of poles of $\bm{C}(\psi)$.

From Eq.~\eqref{eq:A_B_from_G} we can directly obtain:
\begin{equation}\label{eq:A_power}
    \bm{A}_{(u)}(\psi) = \frac{1 - (2u+1)\psi\,\ell^2}{\left(1 - \psi\,\ell^2\right)^{u+1}}\,, \qquad
 \bm{B}_{(u)}(\psi)= \frac{1 + (u-1)\psi\,\ell^2}{\left(1 - \psi\,\ell^2\right)^{u+1}}\,.
\end{equation}
On the other hand, the modified Friedmann equation reads:
\begin{equation}\label{eq:friedmann_power}
    3H^2 \left(1 - H^2 \ell^2\right)^{-u} = \frac{8\pi\rho_0}{a^4}\,,
\end{equation}
Hence, note that \begin{equation}
    a\sim \left(1-\psi\ell^2\right)^{u/4}
\end{equation}
and so 
\begin{equation}
    \bm{C}\sim a^{-4},\qquad \bm{A}\sim \bm{B}
\sim a^{-4(u+1)/u}
\end{equation}
where indeed the degree of divergence of $\bm C$ is the same as the one required by the regularity conditions discussed before, and where both $\bm A$ and $\bm B$ diverge more than $a^{-4}$.

\begin{figure}
    \centering
    \includegraphics[width=0.6\linewidth]{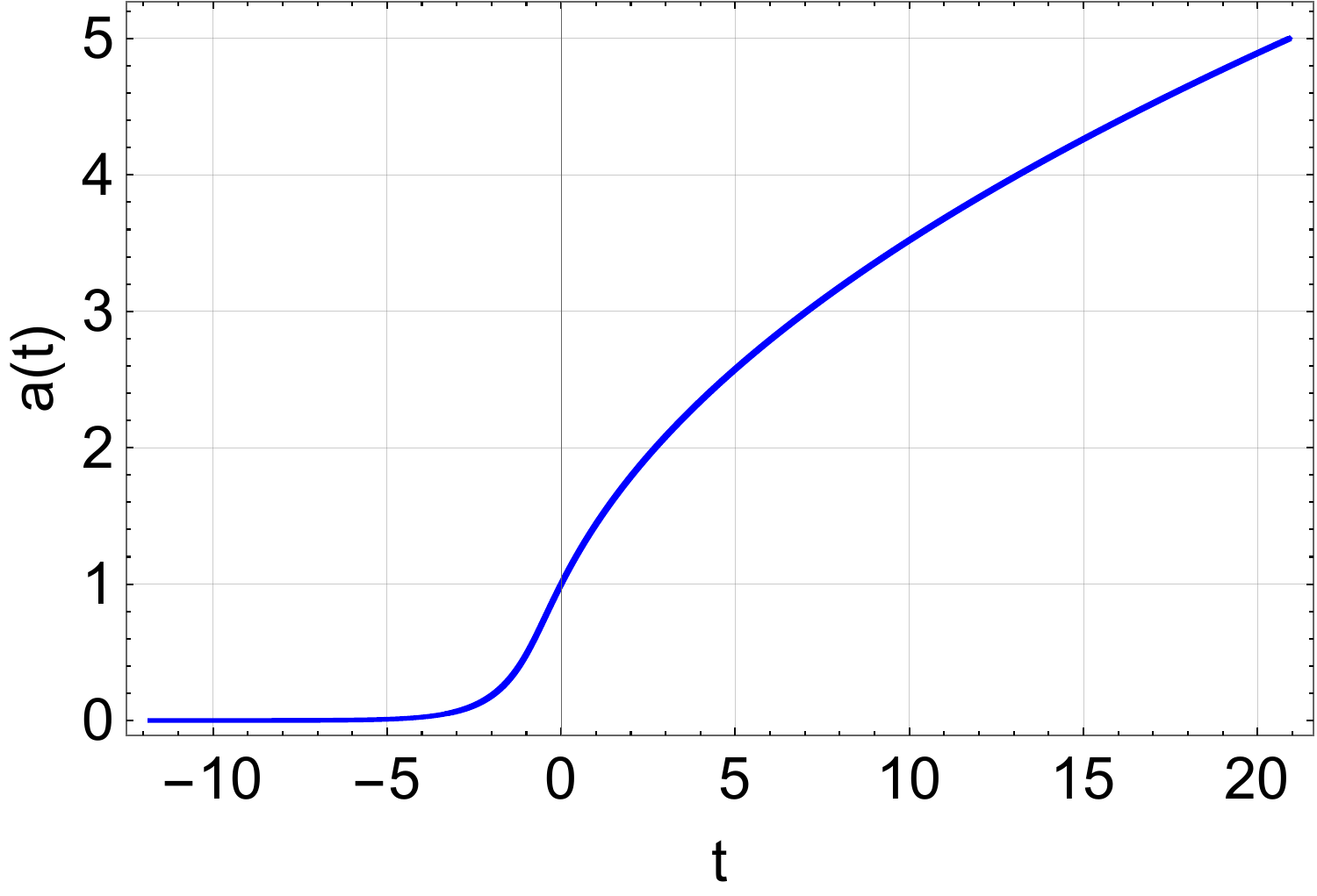}
   
    \caption{Scale factor for a regular cosmological evolution of a flat ($K=0$) universe, obtained as a solution of Eq.~\eqref{eq:friedmann_power} for $n=\ell=\rho_{0}=1$. While not describing a bounce at a finite time, this kind of evolution corresponds to one of the regular cases in the clasification of geodesically complete universes~\cite{Carballo-Rubio:2024rlr,Garcia-Saenz:2024ogr}.}
    \label{fig:regCosmo}
\end{figure}

For $n=1$ we recover the specific theory in the KMT class that also leads to the Hayward regular black hole as a vacuum static solution~\cite{Boyanov:2025pes}. The solution of Eq.~\eqref{eq:friedmann_power} in this case is shown in Fig. \ref{fig:regCosmo} for some specific choices of parameters. Interestingly, this same family can display singular solutions for other matter contents, in particular electromagnetic fields~\cite{Carballo-Rubio:2026mvj}. This shows that the regularity of solutions in the presence of matter fields can, for a given theory, depend on the specific matter content.

\section{Discussion}\label{sec:disc}

In this work we have investigated how the presence of matter fields modifies the conditions for spacetime regularity within a class of gravitational theories characterized by second-order field equations in spherical symmetry. Starting from the general framework of Ref.~\cite{Carballo-Rubio:2025ntd}, in which the Einstein tensor is deformed into a conserved tensor constructed from up to second-order derivatives of the metric, we have systematically derived the constraints that the deformation functions $\bm\alpha$ and $\bm\beta$ must satisfy in order to ensure regularity at the origin across distinct classes of theories coupled to different matter contents. These equations encompass large families of gravitational theories, including non-polynomial gravities in four and higher dimensions~\cite{Colleaux:2017ibe,Colleaux:2019ckh,Bueno:2025zaj,Borissova:2026wmn,Borissova:2026krh}) and quasitopological gravities in dimensions higher than four~\cite{BuenoRegularBlack2025c,Bueno:2024eig,Bueno:2024zsx,Bueno:2025gjg,Hennigar:2025yqm,PinedoSoto:2026hfm,2026arXiv260310110B}, for which our conclusions are relevant.

We first considered spherically symmetric static configurations sourced by a massless scalar field in the Ziprick--Kunstatter class. In general relativity, such configurations lead to naked singularities~\cite{Fisher:1948yn}. We showed that, within the class of theories under consideration, these singularities can be avoided provided the functions $\bm\alpha$ and $\bm\beta$ satisfy specific scaling conditions near $r = 0$, contrary to the situation in vacuum. While these conditions are necessary but not sufficient for global regularity, as they do not rule out singularities at positive values of the radius, we presented a concrete numerical solution that is everywhere regular, demonstrating that global regularity can be realized in practice for this class of theories.
 
We then extended the analysis to time-dependent monopole perturbations of a regular black hole background sourced by a scalar field. Working at linear order in the perturbation amplitude, we derived the constraints on the metric and scalar perturbation functions required to preserve regularity at the origin. The resulting conditions are equivalent to the conditions obtained for the static case. Hence, time-dependence does not introduce additional constraints with respect to the static case, which may be a more general feature for other matter contents. Thereafter, for the other cases studied in the Ziprick--Kunstatter class we focused on the simpler static situations.
 
For electrovacuum solutions, the regularity requirements for static solutions also turn out to be the same as for scalar fields. This matter content has the advantage, with respect to the scalar field case, of admitting analytical static solutions. Thus we presented the analytical regular solution of the same theory in which we found the numerical regular solution in presence of a static scalar field.

We later provided a general analysis of different static matter contents with different behaviors around the center of symmetry to clarify the reason why the results for scalar fields and electromagnetic fields were equivalent. We proved that these two cases are actually particular cases arising from a particular scaling of the matter source in the $r\rightarrow0$ limit. We obtained the most general behavior of perturbative solutions for an arbitrary matter scaling, illustrating that more singular matter sources require stronger modifications of the gravitational dynamics.

Finally we analyzed the Kunstatter--Maeda--Taves family of theories in fully time-dependent but homogeneous and isotropic situations. In particular we chose as a source a perfect fluid of radiation, relevant for cosmological scenarios. We established a no-go theorem for bouncing cosmologies: within the class of theories that admit Minkowski spacetime as a vacuum solution and have second-order equations of motion, no smooth bounce can occur for flat universes. However, we showed that regularity can still be achieved through an emergent de Sitter phase in the asymptotic past, in which the scale factor vanishes only as $t \to -\infty$ while the Hubble parameter remains bounded. We provided explicit examples of theories realizing this mechanism.
 
In conclusion, we have shown that the inclusion of ordinary matter sources does not qualitatively change the capability of second-order spherical gravities to display a less singular behavior than general relativity when characterized in terms of curvature boundedness and geodesic completeness, although these aspects must be carefully analyzed depending on the situation considered.

\acknowledgments

We thank Luis Lehner for valuable discussions. R.C-R. acknowledges financial support provided by the Spanish Government through the Ram\'on y Cajal program (contract RYC2023-045894-I), the Grant No.~PID2023-149018NB-C43 funded~by MCIN/AEI/10.13039/501100011033, and the Severo Ochoa grant CEX2021-001131-S funded by MCIN/AEI/ 10.13039/501100011033. CC is supported by King’s College London through an NMES funded studentship. VV is grateful for the hospitality of Perimeter Institute where part of this work was carried out. Research at Perimeter 
Institute is supported in part by the Government of Canada through the Department of Innovation, Science and 
Economic Development and by the Province of Ontario through the Ministry of Colleges and Universities. This work 
was supported by a grant from the Simons Foundation (1034867, Dittrich).

\appendix

\section{Perturbative expansion of the master perturbation equations near the origin}
\label{app:perturbative}

In this appendix, we present the detailed perturbative analysis of Eqs.~\eqref{eq:pert1} and~\eqref{eq:pert2} near $r=0$, deriving the condition $q\leq -3$ stated in the main text.

\subsection{Setup}

 The relevant near-origin expansions are:
\begin{equation}
f = 1 + f_n\, r^n + \dots\,, \qquad
\bm\alpha \sim \alpha_0\, r^{p}\,, \qquad
\bm\beta \sim \beta_0\, r^{q}\,,
\label{eq:app_background}
\end{equation}
where, as showed in Sec. \ref{sec:vacuum}, the vacuum background equations give $n = -q\geq 2$, $p=-1$, and $\alpha_0 = q\,\beta_0\, f_n$.

For theories of the form~\eqref{eq:alphaZK}, namely $\bm\alpha = (\chi-1)\,\mathrm{d}\bm\beta/\mathrm{d} r$ with $\bm\beta = \bm\beta(r)$, the partial derivatives evaluate to
\begin{equation}
\partial_\chi \bm\alpha = \bm\beta' \sim q\,\beta_0\, r^{q-1}\,, \qquad
\partial_r \bm\alpha = (f-1)\bm\beta'' \sim q(q-1)\,f_n\,\beta_0\, r^{-2}\,, \qquad
\partial_\chi \bm\beta = 0\,.
\label{eq:app_derivatives}
\end{equation}

The scalar field satisfies the Klein--Gordon equation~\eqref{eq:KG} on the regular background. Near $r=0$, using $-\bm\alpha/\bm\beta \sim n\,f_n\,r^{n-1}$ and $2f/r\sim 2/r$, the equation at leading order reduces to
\begin{equation}
\frac{2}{r}\Phi' + \Phi''+ \frac{\omega^2}{4}\Phi= 0.
\end{equation}
Solving this equation perturbatively for the power series expansion of the scalar field in Eq.~\eqref{eq:pertexps} shows that the third term in the equation above is subleading, and results in
\begin{equation}
    s=-1 \qquad \text{or}\qquad s=0.
\end{equation}

\subsection{Scalar branch \texorpdfstring{$s=-1$}{s=.1}}
We here consider the case $s=-1$. Hence,
\begin{equation}
\Phi = \frac{\varphi_{-1}}{r} + \varphi_0 + \varphi_1\, r + \dots\,, \qquad
\Phi' = -\frac{\varphi_{-1}}{r^2} + \varphi_1 + \dots\,.
\label{eq:app_scalar}
\end{equation}
On the other hand, the radial perturbation functions are expanded as in Eq.~\eqref{eq:pertexps}, namely:
\begin{equation}
H_1(r) = \sum_{i=0}^{\infty} \mathcal{H}_i\, r^i\,, \qquad
K_0(r) = \sum_{i=0}^{\infty} \mathcal{K}_i\, r^i\,.
\label{eq:app_pertexp}
\end{equation}
These metric perturbations induce perturbations of the Kreschmann scalar, which can be expanded as
\begin{align}
R_{\mu\nu\rho\sigma}R^{\mu\nu\rho\sigma}=&-\frac{32i\epsilon e^{-i\omega t}}{\omega r^2}\mathcal{H}_0f_3\nonumber\\
&+\frac{8 i \epsilon e^{-i\omega t}}{\omega r}\left[\mathcal{H}_0\left(2 \omega^2 f_2  + 4 f_2^2  + 11 f_4 \right) + 
 6 f_2 \mathcal{H}_2 + 4 i \omega f_2 \mathcal{K}_1\right]+\mathcal{O}(r^0,\epsilon^2). 
\end{align}
Hence, the conditions for regularity of the Kretschmann scalar at order $\epsilon$ are:
\begin{itemize}
\item If $n=2$ and $f_3\neq 0$: $\;\mathcal{H}_0 = 0$, $\;\mathcal{H}_2 = -\tfrac{2}{3}\,i\omega\,\mathcal{K}_1$.
\item If $n=2$ and $f_3 = 0$: $\;\mathcal{H}_2= \left(-2\omega^2\mathcal{H}_0 - 4f_2\,\mathcal{H}_0 - 11\,f_4\,\mathcal{H}_0/f_2 - 4\,i\omega\,\mathcal{K}_1\right)/6$.
\item If $n=3$ or $n=4$: $\;\mathcal{H}_0=0$. 
\item If $n\geq 5$: no conditions on the perturbations.
\end{itemize}
These conditions have been determined kinematically. On the other hand, the coefficients $\mathcal{H}_0$, $\mathcal{H}_2$ and $\mathcal{K}_1$ in the series expansions above are constrained by the dynamical equations~\eqref{eq:pert1} and~\eqref{eq:pert2}. By solving the latter equations perturbatively to obtain the values of $\mathcal{H}_0$, $\mathcal{H}_2$ and $\mathcal{K}_1$ and compare these with the regularity conditions above, we can determine which theories are regular in the presence of these perturbations.

\subsubsection{Expansion of Eq.~\eqref{eq:pert1}}

Omitting the overall prefactor $\epsilon\,e^{-i\omega t}/(4\bm\beta^2 r^3 f)$, Eq.~\eqref{eq:pert1} can be organized as the sum of three terms:
\begin{align}
&\underbrace{\bm\beta^3\!\left(\omega^2 H_1\, r^2 + f^2 J\right)}_{\text{Group A}}
+ \underbrace{\bm\alpha\,\bm\beta\, H_1\, r^2\!\left(\bm\alpha + 2f\,\partial_\chi\bm\alpha\right)}_{\text{Group B}} \nonumber\\[4pt]
&- \underbrace{\bm\beta^2 r\!\left[\bm\alpha\!\left(2i\omega K_0\, r + 4fH_1 + 3frH_1'\right) + fr\!\left(-16i\pi\omega r\,\Phi\Phi' + H_1\,\partial_r\bm\alpha\right)\right]}_{\text{Group C}} = 0\,,
\label{eq:app_master1_groups}
\end{align}
where we have defined $J \equiv -2H_1 + 2rH_1' + r^2 H_1''$.

\paragraph*{Leading order.}
We have $J= -2\mathcal{H}_0+\mathcal{O}(r^1)$. The leading terms in the three groups are:
\begin{center}
\begin{tabular}{lc}
\hline
Group & Leading term  \\
\hline
A & $-2\beta_0^3\,\mathcal{H}_0 \;r^{3q}$ \\
B & $2\alpha_0\,\beta_0\,\mathcal{H}_0 q\,\beta_0 \;r^{2q}$ \\
C & $-(4 \alpha_0\,\mathcal{H}_0+16i\pi\omega\,\varphi_{-1}^2\,)\;\beta_0^2r^{2q}$ \\
\hline
\end{tabular}
\end{center}

Since $3q < 2q$ for $q\leq-2$, terms in group~A are the leading terms for $r\rightarrow0$, implying
\begin{equation}
\mathcal{H}_0 = 0\,.
\label{eq:app_H0}
\end{equation}

\paragraph*{Next order.%: $r^{3q+2}$ vs $r^{2q}$.
}
Setting $\mathcal{H}_0=0$, so that $H_1 = \mathcal{H}_1\,r + \mathcal{H}_2\,r^2+\mathcal{O}(r^3)$, we find $J = 4\mathcal{H}_2\,r^2 + \mathcal{O}(r^3)$. The leading terms for $\mathcal{H}_0=0$ are then:

\begin{center}
\begin{tabular}{lc}
\hline
Group$\qquad$ & Leading term $\left(\mathcal{H}_0=0\right)$  \\
\hline
A & $4\beta_0^3\,\mathcal{H}_2\; r^{3q+2}$\\
B & $ 2\alpha_0\,\beta_0^2\,\mathcal{H}_1 q\;r^{2q+1}$ \\
C & $-16i\pi\omega\,\varphi_{-1}^2\,\beta_0^2\;r^{2q}$ \\
\hline
\end{tabular}
\end{center}

The contribution from group~B is subdominant with respect to the group~C one, while the relative importance of the contributions of groups~A and~C depends on the value of $q$ with respect to the critical threshold $3q+2 = 2q$, i.e., $q=-2$.

For $q< -2$, $3q+2 < 2q$ and the contribution from group~A dominates, imposing $\mathcal{H}_2 = 0$.

For $q=-2$, $3q+2 = 2q = -4$ and the two contributions enter at the same order, giving
\begin{equation}
4\beta_0^3\,\mathcal{H}_2 - 16i\pi\omega\,\varphi_{-1}^2\,\beta_0^2 = 0
\qquad\Longrightarrow\qquad
\mathcal{H}_2 = \frac{4i\pi\omega\,\varphi_{-1}^2}{\beta_0}\,.
\label{eq:app_H2_forced}
\end{equation}

\subsubsection{Expansion of Eq.~\eqref{eq:pert2}}

Omitting the overall prefactor $\epsilon e^{-i \omega t}/\left(2 \omega \bm{\beta}^2 r^3 f^2 \right)$, Eq.~\eqref{eq:pert2} can be organized into three terms:

\begin{align}
&\underbrace{\omega\bm\beta r\!\left[\bm\beta\!\left(\bm\beta(\omega^2 K_0\, r + f^2(4K_0' + rK_0'')) + rf\,K_0\,\partial_r\bm\alpha - 32\pi f^2 r\,\Phi\Phi'\right) + 16\pi r^2 f^2\,\Phi\Phi'\,\partial_\chi\bm\alpha\right]}_{\text{Group D}}\nonumber\\
&+ \underbrace{\bm\alpha f\bm\beta\left[\bm\beta(-2if(2H_1 + rH_1') + \omega r(5K_0 + rK_0')) - 2\omega r^2 \left( 8 \pi r \Phi \Phi' +K_0\,\partial_\chi\bm\alpha\right)\right]}_{\text{Group E}} \nonumber\\[4pt]
&- \underbrace{\bm\alpha^2 r\,\bm\beta\,(2ifH_1 + \omega K_0\, r)}_{\text{Group F}} = 0\,,
\label{eq:app_master2_lines}
\end{align}

\paragraph*{Leading order.}
With $\mathcal{H}_0 = 0$ ($H_1 = \mathcal{H}_1\,r + \dots$) and $K_0 = \mathcal{K}_0 + \mathcal{K}_1\,r + \dots$:

\begin{center}
\begin{tabular}{lc}
\hline
Group & Leading term(s) $\left(\mathcal{H}_0=0\right)$\\
\hline
D  & $4\omega\,\beta_0^3\,\mathcal{K}_1 \; r^{3q+1}+16\pi\omega\,(2-q)\,\varphi_{-1}^2\,\beta_0^2\;r^{2q-1}$ \\
E & $\alpha_0\,\beta_0^2\,\left[-6i\mathcal{H}_1 + \omega\mathcal{K}_0(5-2q)\right]\;r^{2q}$ \\
F & $\alpha_0^2\,\beta_0\,(2i\mathcal{H}_1 + \omega\mathcal{K}_0)\;r^{q}$ \\
\hline
\end{tabular}
\end{center}

The two competing leading powers are $r^{3q+1}$ and $r^{2q-1}$ in group D, with the threshold at $3q+1 = 2q-1$, again $q=-2$.

For $q\leq -3$, $3q+1 < 2q-1$ and the geometric terms in group D dominate, resulting in:
\begin{equation}
\mathcal{K}_1 = 0\,.
\label{eq:app_K1}
\end{equation}

For $q=-2$, $3q+1 = 2q-1 = -5$ and the two terms in group D balance:
\begin{equation}
4\omega\,\beta_0^3\,\mathcal{K}_1 + 16\pi\omega\,(2-q)\,\varphi_{-1}^2\,\beta_0^2 = 0
\qquad\Longrightarrow\qquad
\mathcal{K}_1 = -\frac{16\pi\,\varphi_{-1}^2}{\beta_0} \neq 0\,.
\label{eq:app_K1_forced}
\end{equation}

Hence, for $q=-2$ the relation $\mathcal{H}_2=-i\mathcal{K}_1/4$ follows.

\subsubsection{Overview}

For $q=-2$ ($n = -q = 2$), the perturbation equations force $\mathcal{H}_0=0$ (Eq.~\eqref{eq:app_H0}), $\mathcal{H}_2 \neq 0$ (Eq.~\eqref{eq:app_H2_forced}) and $\mathcal{K}_1\neq 0$ (Eq.~\eqref{eq:app_K1_forced}). For $\mathcal{H}_0=0$, the regularity conditions of the Kretschmann scalar at order $\epsilon$ for $n=2$ become (regardless of the value of $f_3$):
\begin{equation}
\mathcal{H}_0 = 0\,, \qquad \mathcal{H}_2 = -\tfrac{2}{3}\,i\omega\,\mathcal{K}_1\,.
\label{eq:app_Kretschmann_n2}
\end{equation}
The second relation is incompatible with the perturbative solutions of Eqs.~\eqref{eq:pert1} and~\eqref{eq:pert2} for $q=-2$ discussed above.

For $q\leq -3$ ($n\geq 3$), the scalar source is subleading: the geometric terms proportional to $\bm\beta^3$ dominate, imposing $\mathcal{H}_0 = \mathcal{H}_2 =\mathcal{K}_1 = 0$ from the leading orders alone. For $n=3$, the only Kretschmann condition is $\mathcal{H}_0=0$, which is automatically satisfied by the solutions of the master equations. For $n\geq 4$, no conditions are imposed on the perturbations. In both cases, the perturbative system is consistent. Going further in the perturbative order is possible to find consistent relations between the remaining coefficients of the expansion. In particular every coefficient $\mathcal{H}_i$, $\mathcal{K}_i$ can be written as a non-trivial linear combination of the three leading ones $\mathcal{H}_1$, $\mathcal{K}_0$, $\varphi_{-1}$ and of the metric coefficients.

Therefore, regularity of curvature invariants in the presence of scalar perturbations requires
\begin{equation}
q \leq -3\,
\label{eq:app_final}
\end{equation}
confirming that $\bm\beta$ must diverge at least as $r^{-3}$ near the origin. This is more restrictive than the vacuum condition $q\leq -2$, demonstrating that the presence of scalar perturbations further constrains the class of viable theories.

\subsection{Scalar branch \texorpdfstring{$s=0$}{s=0}}

As noted before, the Klein Gordon equation~\eqref{eq:KG} near $r=0$ admits two roots: the irregular branch $s=-1$, analyzed above, and the regular branch $s=0$. For completeness, here we repeat the analysis for the latter, which yields:
\begin{equation}
\Phi = \varphi_0 + \varphi_1\, r + \varphi_2\, r^2+ \dots\,, \qquad \Phi' = \varphi_1 + 2\,\varphi_2\, r+..,
\end{equation}
so that $\Phi\Phi' = \mathcal{O}(r^0)$, three powers of $r$ less singular than in the $s=-1$ case where $\Phi\Phi'\sim -\varphi_{-1}^2\,r^{-3}$.

\subsubsection{Expansion of Eq.~\eqref{eq:pert1}}
At leading order, Group A still gives $-2\beta_0^3\,\mathcal{H}_0\,r^{3q}$ and dominates over all other terms since $3q<2q$ for $q\leq -2$, imposing \begin{equation}
    \mathcal{H}_0=0
\end{equation} 
as before. Setting $\mathcal{H}_0=0$ and considering the next to leading order, the dominant term of the scalar source in Group C now scales as $r^{2q+1}$, so the threshold between Groups A and C shifts to $3q+2=2q+1$, i.e.\ $q=-1$. Since $q\leq -2$ throughout, Group A dominates for all allowed values of $q$, giving
\begin{equation}
\mathcal{H}_2 = 0 \qquad \text{for all } q\leq -2\,.
\end{equation}
No balancing between geometric and scalar terms occurs at any order.

\subsubsection{Expansion of Eq.~\eqref{eq:pert2}}
The scalar source in Group D now scales as $r^{2q+2}$. The threshold with the geometric contribution $4\omega\,\beta_0^3\,\mathcal{K}_1\,r^{3q+1}$ shifts to $3q+1=2q$, i.e.\ $q=-1$. Since $q\leq -2$, the geometric term in Group D dominates for all allowed $q$, imposing
\begin{equation}
\mathcal{K}_1 = 0 \qquad \text{for all } q\leq -2\,.
\end{equation}
\subsubsection{Kretschmann conditions}
With $\mathcal{H}_0=\mathcal{H}_2=\mathcal{K}_1=0$ enforced by the dynamics, the Kretschmann regularity condition is automatically satisfied. Hence the system is consistent for all $q\leq -2$, and the vacuum regularity bound is not further restricted by the presence of $s=0$ scalar perturbations.
\subsubsection{Overview}
The $s=0$ case contrasts with the $s=-1$ case, where the $r^{-3}$ singularity of $\Phi\Phi'$ forces a non-trivial competition between scalar and geometric terms at $q=-2$, leading to values of $\mathcal{H}_2$ and $\mathcal{K}_1$ that are incompatible with Kretschmann regularity. The additional restriction $q\leq -3$ derived above is therefore entirely a consequence of selecting the branch $s=-1$, and is absent if the branch $s=0$ is chosen instead. However, there is no reason for excluding the scalar field solution with $s=-1$ from the analysis. Therefore, the condition that the theory should meet to ensure regularity even with scalar field perturbations is the more restrictive one, given by the $s=-1$ case.

\section{Cosmologies with spatial curvature}\label{sec:K_nonzero}
 
When spatial curvature is included, the modified Friedmann equation reads
\begin{equation}\label{eq:friedmann_K}
    \psi\,\bm{C}(\psi) = \frac{8\pi\rho_0}{a^4}\,, \qquad
    \psi = \frac{K + \dot{a}^2}{a^2}\,.
\end{equation}
At a turning point ($\dot{a} = 0$), $\psi = K/a^2$, which vanishes only for $K = 0$.
This is the origin of the qualitative difference between flat and curved cosmologies.
 
For $K = 0$, the left-hand side of Eq.~\eqref{eq:friedmann_K} vanishes at $\dot{a} = 0$
while the right-hand side remains strictly positive for $\rho_0 > 0$. For $K = +1$, however, $\psi = 1/a^2 > 0$ at the turning point,
and Eq.~\eqref{eq:friedmann_K} becomes
\begin{equation}\label{eq:bounce_cond}
    \bm{C}\!\left(\frac{1}{a_0^2}\right) = \frac{8\pi\rho_0}{a_0^2}\,,
\end{equation}
which is a well-defined algebraic equation for $a_0$ that generically admits solutions.
In fact, general relativity itself ($\bm{C} = 3$) has a solution, namely $a_0 = (8\pi\rho_0/3)^{1/2}$.
 
For $K = -1$, the turning point gives $\psi = -1/a_0^2 < 0$, and the left-hand side of
Eq.~\eqref{eq:friedmann_K} is negative for any $\bm{C} > 0$. Since the right-hand
side is positive, no turning point exists for physically reasonable theories. The open case
is therefore analogous to the flat one, as the singularity can't be resolved through a bounce. 
 
\subsubsection{Distinguishing bounce from turnaround}
 
Not every turning point is a bounce. A true bounce (local minimum of $a$) requires
$\ddot{a} > 0$, while a turnaround (local maximum) has $\ddot{a} < 0$.
Writing $\dot{a}^2 = F(a) \equiv \psi(a)\,a^2 - K$, the acceleration at a turning point is
$\ddot{a} = F'(a_0)/2$. Differentiating $F$ and using the constraint equation to
eliminate $\psi'(a)$ yields, after simplification,
\begin{equation}
    \ddot{a}\big|_{a_0} = \frac{1}{a_0}
    \frac{\psi_0\,\bm{C}'(\psi_0) - \bm{C}(\psi_0)}
    {\bm{C}(\psi_0) + \psi_0\,\bm{C}'(\psi_0)}\,,
\end{equation}
where $\psi_0 = 1/a_0^2$. Thus $\ddot{a}$ is positive either when numerator and denominator are both positive
(\emph{case~I}: $\psi_0\,\bm{C}' > \bm{C}$ with $f'(\psi_0) > 0$, where
$f(\psi) \equiv \psi\,\bm{C}(\psi)$ and $f' = \bm{C} + \psi\bm{C}'$),
or both negative (\emph{case~II}: $f'(\psi_0) < 0$).
We will show now that the \emph{case~II} does not correspond to a physically viable bounce, because any
trajectory connecting it to the low-curvature regime must cross a curvature singularity. 

The low-curvature regime is characterized by
$f'(0) = \bm{C}(0) = 3 > 0$, as required by the Newtonian limit. A \emph{case~II} bounce,
by definition, sits at a point where $f'(\psi_0) < 0$. Since $f'(\psi)$ is continuous,
any cosmological trajectory that starts in the low-curvature regime (where $f' > 0$)
and reaches the \emph{case~II} turning point (where $f' < 0$) must pass through an intermediate
value $\psi_* $ at which
\begin{equation}
    f'(\psi_*) = \left[\psi\,\bm{C}(\psi)\right]'\Big|_{\psi_*} = 0\,.
\end{equation}
In other words, the function $f(\psi)$ is non-monotonic: it grows from $f(0)=0$, reaches
a local maximum $f_\mathrm{max} = f(\psi_*)$, and decreases beyond it. The \emph{case~II}
branch lies entirely on the decreasing side.

Taking a time derivative of the constraint
$f(\psi) = 8\pi\rho_0/a^4$ and using
$\dot{\psi} = 2H(\ddot{a}/a - \psi)$, valid for any $K$, yields the exact acceleration
equation
\begin{equation}\label{eq:accel_general}
    \frac{\ddot{a}}{a} = \psi - \frac{2 f(\psi)}{f'(\psi)}\,.
\end{equation}

Equation~\eqref{eq:accel_general} makes the pathology manifest: as the trajectory
approaches $\psi \to \psi_*$, the numerator $f(\psi_*) = f_\mathrm{max} > 0$ remains
finite while the denominator $f'(\psi) \to 0$, so that
\begin{equation}
    \ddot{a} \;\longrightarrow\; \mp\infty\,.
\end{equation}
 Correspondingly, the Ricci scalar
\begin{equation}
    R = 6\left(\frac{\ddot{a}}{a} + \frac{\dot{a}^2 + K}{a^2}\right)
    \;\longrightarrow\; \infty
\end{equation}
diverges at finite $a$ and finite proper time.

Therefore, $\ddot{a} > 0$
if and only if (\emph{case~I}):
\begin{equation}\label{eq:bounce_min}
\frac{d\ln\bm{C}}{d\ln\psi}\bigg|_{\psi_0} > 1.
\end{equation}
 In general relativity
$\bm{C} = 3$ is constant, so the left-hand side vanishes identically, every turning
point is a turnaround (maximum), never a bounce. A bounce requires $\bm{C}$ to grow
sufficiently fast, which is naturally achieved by theories with a pole at some
$\psi_c > 0$.
 
For the power-law family $\bm{C}_n = 3/(1-\psi\ell^2)^n$,
Eq.~\eqref{eq:bounce_min} reduces to $\psi_0\ell^2 > 1/(n+1)$, or equivalently
$a_0 < \sqrt{n+1}\,\ell$. The turning point closer to $a = 0$ (small $a_0$, large
$\psi_0$) is a bounce, while the one at large $a_0$ is a GR-like turnaround.
When both turning points exist, the universe oscillates between
$a_\mathrm{min} \sim \ell$ (bounce) and $a_\mathrm{max} \sim (8\pi\rho_0/3)^{1/2}$
(turnaround), remaining regular for all time.

\noindent The presence of a pole in $\bm{C}$ resolves the singularity in all three
cases, but through qualitatively different mechanisms: a bounce for $K = +1$ (where
spatial curvature provides the nonzero $\psi$ at the turning point) and an emergent
de~Sitter phase for $K = 0$ and $K = -1$ (where the pole caps $H$ at $1/\ell$).

\bibliographystyle{utphys}

\bibliography{refs}

\end{document}